\pdfoutput=1
\newcommand*{\ATLASLATEXPATH}{}
\documentclass[atlasdraft=true, texlive=2016, UKenglish,cernpreprint,]{\ATLASLATEXPATH atlasdoc}
 
\usepackage{\ATLASLATEXPATH atlaspackage}
\usepackage{\ATLASLATEXPATH atlasbiblatex}
 
\usepackage{\ATLASLATEXPATH atlasphysics}
 
\graphicspath{{logos/}{pdf/}}
 
\usepackage{MDET-2018-01-PAPER-defs}
 
\usepackage{lineno}
\usepackage{amsmath}
\usepackage{bm}
\usepackage{\ATLASLATEXPATH atlasunit}
\usepackage{\ATLASLATEXPATH atlasmisc}
\usepackage{float}
\AtlasTitle{Resolution of the ATLAS muon spectrometer
monitored drift tubes in LHC Run~2}

\AtlasAbstract{
The momentum measurement capability of the ATLAS muon spectrometer
relies fundamentally on the intrinsic single-hit spatial resolution of the
monitored drift tube precision tracking chambers. Optimal resolution
is achieved with a dedicated calibration program that addresses the
specific operating conditions of the 354\,000 high-pressure drift tubes
in the spectrometer.
The calibrations consist of a set of timing offsets and drift time to drift
distance transfer relations, and result in chamber resolution functions.
This paper describes novel algorithms to obtain precision calibrations
from data collected by ATLAS in LHC Run~2 and from a gas monitoring chamber,
deployed in a dedicated gas facility.
 
The algorithm output consists of a pair of correction constants per
chamber which are applied to baseline calibrations, and which have been determined
for ATLAS Run~2. The final single-hit spatial resolution,
averaged over 1172 monitored drift tube chambers, is $81.7 \pm 2.2~\upmu$m.}
 
\author{The ATLAS Collaboration}
 
\AtlasRefCode{MDET-2018-01}
 
\PreprintIdNumber{CERN-EP-2019-091}
 
\AtlasDate{\today}

\arXivId{1906.12226}

\AtlasJournal{JINST}
\AtlasJournalRef{JINST 14 (2019) P09011}
\AtlasDOI{10.1088/1748-0221/14/09/P09011}

\hypersetup{pdftitle={Resolution of the ATLAS Muon Spectrometer
Monitored Drift Tubes in the LHC Run~2},pdfauthor={The ATLAS Collaboration}}
 
\begin{document}
 
\maketitle
 
\tableofcontents
 
\section{Introduction}
\label{sec:intro}
 
This paper summarizes the calibration procedure and reports the single-hit
spatial resolution of the monitored drift tubes (MDT) in
the ATLAS muon spectrometer (MS) during Run~2 of the Large
Hadron Collider (LHC), spanning the 2015-2018 operating period.
In this context, \textit {single-hit} refers to the
signal produced by a charged particle passing through a drift tube.
The calibrations use data from both the spectrometer MDT chambers and
a dedicated gas monitor MDT chamber (GMC), and are performed at the ATLAS
muon calibration centers~\cite{Dai_2010}.
\par
The MS design momentum resolution $\delta p/p$ is $\sim$3\% at 100~\GeV\
and around 10\% at 1~\TeV~\cite{CERN-LHCC-97-022}, allowing both
Standard Model measurements and new physics searches in
processes containing high-momentum final-state muons~\cite{EXOT-2016-05}.
Such searches may include the decay of heavy resonances into
high transverse momentum (\PT) muons, demanding superior spectrometer
performance for their discovery.
For muons of \PT\ above 100~\GeV,
the drift tube single-hit resolution (and to a lesser extent, the
chamber alignment) becomes an important contribution to the transverse
momentum resolution, and is the dominant factor above
250~\GeV~\cite{CERN-LHCC-97-022}.
At high momentum, the \PT\ resolution $\sigma_{\pT\ } $ is determined by the
relation $\sigma_{\pT\ } / \pT\ \propto
\mathrm{\sigma_{res}} \times \pT\ $~\cite{PERF-2011-01},
where $\mathrm{\sigma_{res}}$
is the single-hit spatial resolution. The  design resolution is
$\mathrm{\sigma_{res}} = 80~\upmu$m~\cite{CERN-LHCC-97-022}.
\par
The  objective of the MDT calibration program is to achieve this
resolution for all 354\,000 drift tubes assembled into 1172 MDT chambers
throughout all ATLAS data taking in LHC Run~2.
The primary MDT calibration ingredients consist of tube \TZ~timing offsets
(referred to here as \TZ offsets or just \TZ) and chamber-specific drift electron  time to drift distance
transfer relations, hereafter referred to as \RT~functions.
The \TZ~offset is primarily a function of cable length and electronics delays,
while the \RT~function depends on the gas properties.
\par
The operating conditions of the MDT detectors in the ATLAS cavern are
characterized by large temperature gradients, occasional component faults,
variations in humidity, background radiation, flow, pressure and mixture
instabilities of the gas supply, all of which affect the timing response.
In order to achieve the optimal resolution the calibration constants are
routinely  monitored, corrected for many of these second-order effects,
and updated.
\par
This paper describes the algorithms which generate a pair of correction
parameters for the \TZ~offsets and the \RT~functions.
These pairs of chamber-specific constants compensate for a number of
systematic uncertainties associated with the operating conditions of the MDT
chambers.
As a final result, the MDT tube spatial resolution that has been achieved
in ATLAS Run~2 is presented. Unless otherwise noted, all reported errors are
statistical.

\section{ATLAS detector}
\label{MSdescription}
 
The ATLAS experiment~\cite{PERF-2007-01} at the LHC is a multipurpose
particle detector with a forward--backward symmetric cylindrical geometry
and a near $ 4 \pi $ coverage in solid angle.\footnote{ATLAS
uses a right-handed coordinate system with its origin at
the nominal interaction point  in the center of the detector and the
$z$-axis along the beam pipe.
The $x$-axis points from the interaction point to the center of the LHC ring,
and the $y$-axis points upwards.
Cylindrical coordinates ($r$, $\varphi$) are used in the transverse plane,
$\varphi$  being the azimuthal angle around the $z$-axis.
The pseudorapidity is defined in terms of the polar angle $\theta$ as
$\eta = - \ln \tan(\theta/2)$.
The momentum in the transverse plane is denoted by \PT.}
It consists of an inner tracking detector (ID) surrounded by a thin
superconducting solenoid providing a 2~T axial magnetic field,
electromagnetic and hadronic calorimeters, and a muon spectrometer.
The ID covers the pseudorapidity
range $ |\eta| < 2.5$.
It consists of silicon pixel, silicon microstrip, and transition-radiation
tracking detectors. Lead/liquid-argon (LAr) sampling calorimeters provide
electromagnetic (EM) energy measurements with high granularity.
A steel/scintillator-tile hadronic calorimeter covers the central
pseudorapidity range ($|\eta| < 1.7$).
The endcap and forward regions are instrumented with LAr calorimeters for
EM and hadronic energy measurements up to $|\eta| = 4.9$.
 
\par
The ATLAS muon spectrometer is cylindrical, 22~m in diameter and 45~m
in length, surrounding
the calorimeters with an acceptance coverage of $ 2\pi $ in azimuth and
${}\pm$2.7 in pseudorapidity.
It consists of a barrel and two endcap sections located in an air-core
toroidal magnetic field generated by eight coils for each section.
The field integral of the toroids ranges between 2.0 and 6.0~T\,m for most
of the acceptance, while the field magnitude at the MDT chambers
ranges from near zero to 0.2~T in most of the endcap region ($1.05<|\eta|<2.7 $),
and averages about 0.6~T in the barrel ($ |\eta| < 1.05 $)
region~\cite{CERN-LHCC-97-022}.
The MS identifies muons, provides muon triggers and, integrating with
the inner detector information, measures their charge sign and momenta.
The MDT chambers perform precision charged-particle tracking in the plane
defined by the beam axis ($z$) and the radial distance to the beam ($r$).
The momentum measurement is based on the track curvature in these
$r$--$z$ coordinates, while resistive-plate
chambers (RPC) and thin-gap chambers (TGC) provide triggering in the barrel
and endcaps respectively, and readout of the so called \textit{second coordinate}
in the $r$--$\varphi$ non-bending plane. The MDT chambers are assembled in three nested cylindrical barrel stations
and three main coaxial discs in each of the endcap regions. A fourth annulus of endcap chambers covers a limited pseudorapidity
region spanning from the barrel to the endcap. They
are constructed of pairs of close-packed multilayers of 3~cm diameter
cylindrical aluminum drift tubes, 1~m to 6~m long for the inner stations
close to the beam and outer stations, respectively. 
Each multilayer comprises either three or four single planes, depending on the
barrel (endcap) chamber radial ($z$-axis) position, with each plane containing from 12 to 72 tubes.
A $ 50 \; \upmu$m diameter anode wire is centered to within $30 \; \upmu$m of
the axis of a 1.46~cm (1.5~cm) inner (outer) radius aluminum tube.
The MDT gas is a mixture of $93\%$ Ar and $7\%$ CO$_2^{}$ plus a few
hundred ppm of water vapour at 3~bar pressure.  The gas flows in a common trunk line to
distribution racks and is fed in parallel to all drift tubes at a rate of roughly
one tube volume per day.
The total gas throughput to the entire MDT system is about $ 2 \times
10^6~\mathrm{L/day}$. The tubes are operated at 3080~V
anode voltage to deliver a gas avalanche amplification gain of $20\,000$.
The single-hit efficiency for a minimum-ionizing particle is nearly
100\%~\cite{MUON-2010-01}.
\par
First-level triggers~\cite{TRIG-2016-01}   are implemented in hardware and combine muon spectrometer
and calorimeter measurements to accept events at a maximum rate of 100~kHz.
These first-level triggers are followed by more refined software-based
triggers that reduce  the accepted event rate to 1~kHz on average.
 
\section{Analysed data}\label{DataSelection}
The results presented here are based on a subset of the ATLAS Run~2
proton--proton collision data at a center-of-mass energy $\sqrt{s}=13$~\TeV.
Specifically, this analysis uses  $\mathrm{ 1.3 \; fb^{-1}}$ of data that were acquired in 2015,
$\mathrm{ 23.8 \; fb^{-1}}$ in 2016, and $\mathrm{ 22.3 \; fb^{-1}}$ in 2017.
The 2015 and 2017 data include $\mathrm{ 7.9 \; pb^{-1} }$
and $\mathrm{ 55.3 \; pb^{-1} }$, respectively, collected with the MS toroid
magnet turned off.
The data set is from the primary ATLAS physics data stream that includes all
triggers and was processed with data quality selections common to all physics
analyses.
The selected set of data runs is distributed along the whole data acquisition
period for each year, allowing checks of the stability of the parameters
under study in different conditions. Each single run contains
enough muon tracks per chamber to allow the determination of the \RT~and resolution functions.
Two corrections that depend on the non-bending plane second coordinate are applied
to the time of the MDT signal by the offline event reconstruction software:
(1) the magnetic-field--induced Lorentz angle effect on the drift
time~\cite{Horvat} and (2) the signal propagation time
associated with the hit position along the wire.
\par
This paper uses single-hit information from muon tracks with \PT ${}> 20$~\GeV\
that are reconstructed in the ID and MS.
The data include drift times, drift radii and timing offsets as
used by the offline tracking reconstruction algorithms.
A minimum of 1000 tube hits per chamber per run is required to obtain
stable fit results in the analysis.
As a consequence of these selection criteria the number of chambers for
which results are obtained has a run-to-run variation of 0.3\%.
When possible, multiple consecutive runs with the same gas conditions are
grouped together in order to reach a minimum number of muon tracks per chamber.
 
\section {MDT calibration}
This section introduces the drift radius measurement and the basic calibration
parameters necessary for its determination.
The operational definition of the single-hit spatial resolution as applied in
this analysis is then presented.
 
\subsection{Drift radius determination}
 
The MDT drift tube signal, or hit,  provides the pulse arrival time, and a
quantity reconstructed as the collected electric charge of the signal pulse.
Read-out is accomplished with 24-channel front-end cards.
These cards amplify, shape and discriminate the signals with programmable
thresholds, then digitize the data.
The hit time data are recorded in  $1.2~\upmu$s windows initiated by a first-level
trigger signal with a 0.78125~ns minimum digitization time.
The charge contained in the leading edge of the signal is stored as voltage on a
capacitor and read out as a run-down time using the Wilkinson
technique~\cite{MDTelx:2008}.
\par
The signal arrival time is relative
to a specific proton--proton interaction time called the bunch crossing
identifier. The bunch crossing time is measured in 25~ns ticks
synchronized with an LHC global 40~MHz clock, and is subtracted in the read-out
electronics hardware to yield the raw pulse arrival time.
This pulse arrival time includes, in addition to the drift time, several timing
offsets:
the time-of-flight of the particle (muon) from the proton--proton interaction
point to the tube, the signal propagation time along the tube wire,
plus various fixed electronics and cable latencies.
The sum of all these is referred to as the \TZ~offset.
The \TZ~offset therefore represents the drift time produced by a particle crossing
at the tube wire.
\par
\begin{figure}[htpb]
\begin{minipage}[t]{0.48\linewidth}
\centering
\vspace{0pt}
\captionsetup{width=.95\linewidth}
\includegraphics[width=1.0\textwidth]{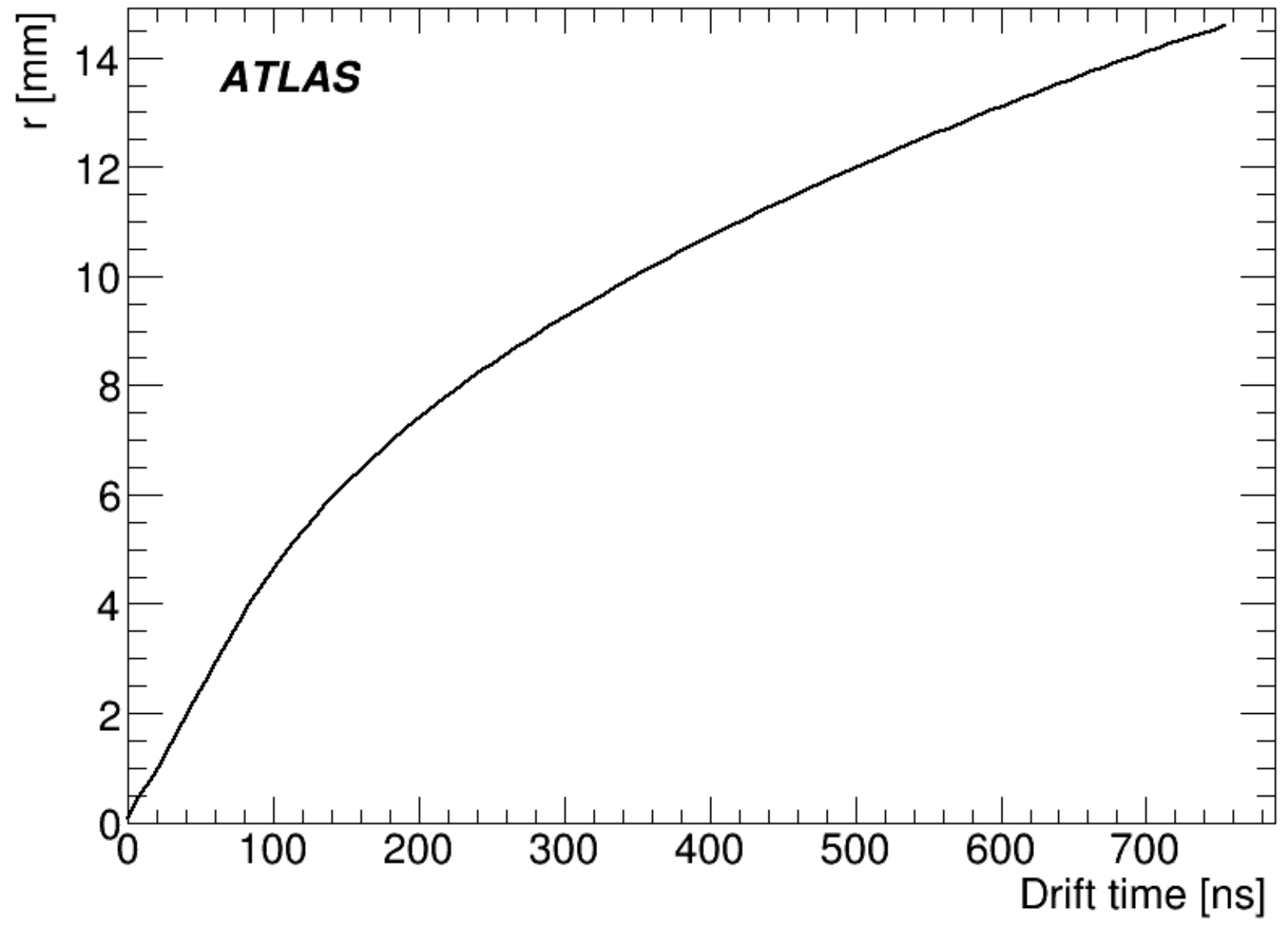}
\caption{Gas monitor chamber \RT~function for 93\%~Ar--7\%~CO$_2^{}$
(and less than 0.1\% of water vapour)
for standard condition: 293~K, 3080~V and 3~bar,
obtained via autocalibration.} 
\label{fig:rtfunction}
\end{minipage}
\hfill
\begin{minipage}[t]{0.48\linewidth}
\centering
\vspace{0pt}
\captionsetup{width=.95\linewidth}
\includegraphics[width=1.0\textwidth]{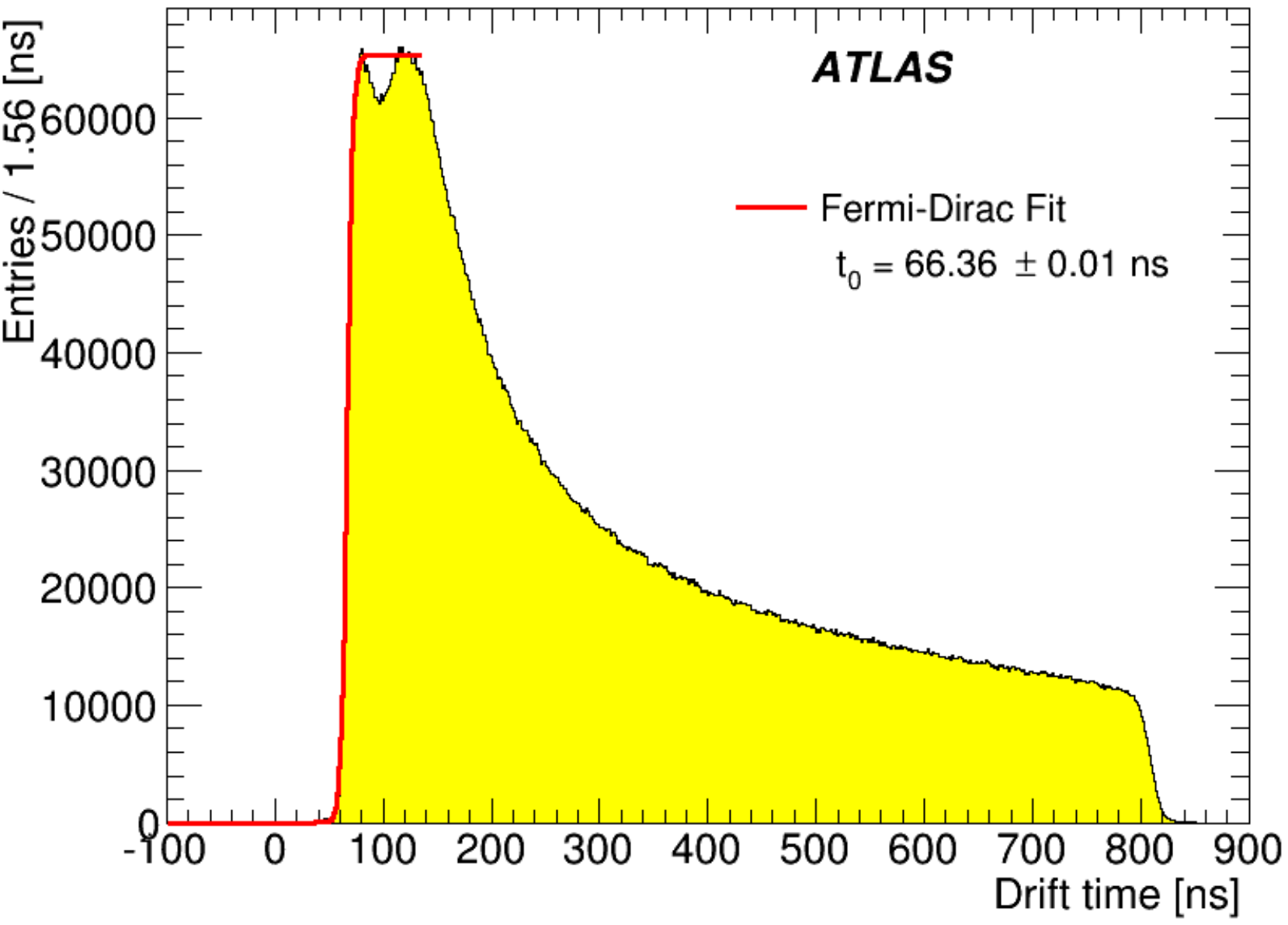}
\caption{Measured drift time spectrum for a generic MDT chamber
with a Fermi--Dirac function fit to the rising edge
of the distribution.}
\label{fig:TDCfit}
\end{minipage}
\end{figure}

The raw time for every hit is converted into a drift time by subtracting
the tube time offset.
The drift time is a measure of the transit time of electrons generated by the
primary ionization and subsequent avalanche multiplication to arrive at the
anode wire.
The resultant drift time is then mapped to the drift distance (circle)  by means
of a time-to-distance~\RT~transfer function, an example
of which is shown in Figure~\ref{fig:rtfunction}.
The best straight line tangent to all such drift circles in a single chamber
forms a track segment.
\par
 
In ATLAS, the \TZ~offset is a property of a single tube, whereas an \RT~function
is associated to a single MDT chamber.
This association is valid to the extent that the operating conditions are uniform.
Since chambers share a common high-voltage supply, gas pressure and thermal
environment, these conditions are generally met at the chamber level.
These \TZ~offsets and \RT~functions, globally called calibration constants,
are generated by the ATLAS calibration computing centers.
\subsection{\TZ~determination}\label{T0determination}
While the \TZ~offset represents the drift time of a hit at the wire,
operationally it is defined as the point of 50\% of maximum amplitude of a
Fermi--Dirac sigmoid function fit to the rising edge of an MDT drift
time spectrum~\cite{Levin:2008zza}, as illustrated in Figure~\ref{fig:TDCfit}.
\TZ~offsets are nominally determined for all tubes in the MS using single tube spectra or, when hit statistics are poor,
the combined spectra  from groups of 24 equal-length tubes sharing a common readout card.
This \TZ~offset definition is subject to small systematic uncertainties of order 1~ns.
These arise from different data selection, different  background level
and operating conditions in the GMC (see Section~\ref{sec:GMC}) and the MDT
systems, which affect the number of hits near the wire, the slope of the
rising edge, and the subsequent fit result. These
systematic effects are relevant for the GMC-based universal \RT\  function described in Sec.~\ref{sec:RT}
which is computed using the GMC \TZ~offsets. They are resolved using a correction parameter,
\DTZ , described in Section~\ref{sec:T0correction}.
 
\subsection{\RT~function determination}
\label{sec:RT}
A standard method to extract the \RT~function is by an autocalibration
algorithm~\cite{AUTOCAL}. This algorithm, using a given set of \TZ~offsets, iterates the linear fit to the drift radii in a chamber,
each time updating the \RT~function until the residuals converge to a minimum.
An autocalibration \RT~function is self-adapting to the chamber conditions,
but a daily implementation of this procedure for every single MDT chamber
requires significant computing resources and a large amount
of data transfer and storage.
\par
The alternative \RT~function calibration method used here adopts the
temperature-corrected universal \RT\ (URT), where the URT is the GMC
\RT~function derived using autocalibration for standard operating conditions
(defined as gas pressure 3~bar, high voltage 3080~V, temperature 293~K, and GMC \TZ~offsets).
The URT method assumes that the absolute measured values of the chamber
operating conditions 
of both the GMC and the MDT chambers are all known and that their
differences can be corrected.
The main sources of systematic differences resulting from this assumption are:
\begin{itemize}
\item The MDT chamber pressures can differ from the 3~bar nominal point
due to several factors: local gas flow rates, pressure gauge location in
the gas flow circuit, and pressure sensor offsets. Furthermore, sensors
used in the GMC and in the MDT gas racks have systematic uncertainties
of 10--20~mbar.
\item The internal MDT gas temperature may be systematically misrepresented by
the thermal sensors. There are typically  eight sensors mounted on the tube
walls at different locations in the chamber.
These sensors are read out periodically, and a daily average value over
all sensors is computed.
However, due to possible thermal gradients across the chambers
(on average $\mathrm{1.8 \pm 0.7 }$~(RMS)~K), and because
the sensors are external to the tubes, this average temperature may not
accurately represent the internal temperature everywhere in the gas
volume.
\item The actual high voltage supplied to chambers might deviate slightly
(by about 1 V) from the nominal 3080~V operating voltage, although
this has an effect of less than 0.1\% on the maximum drift time.
A larger high-voltage deviation generally
indicates equipment failure, not a systematic error, and it is addressed
during the detector access periods.
\end{itemize}
To correct for these effects a single catch-all parameter,
referred to as effective pressure, \DP, is defined.
This parameter is used in a  phenomenological correction function
described in Section~\ref{sec:dpcorrection}.
\par
In summary, an MDT chamber requires two types of calibrations: an
\RT~function and a set of \TZ~offsets, each with a specific set of uncertainties.
Two similar algorithms provide correction factors that compensate for
these uncertainties and lead to improved single-hit spatial resolution.

\subsection{Gas monitor chamber}
\label{sec:GMC}
The GMC~\cite{Levin:2008zza} is an independent MDT chamber that operates in
the surface gas facility, outside of the ATLAS experimental cavern.
It has two gas partitions each independently sampling the common MDT trunk
supply and exhaust lines that serve the entire MDT system, and records
cosmic-ray muons at a rate of about 14~Hz.
It is housed in a stable thermal enclosure where all operating parameters,
gas flow, pressure, temperature, and high voltage,
are continuously monitored.
\par
\begin{figure}[htpb]
\begin{minipage}[t]{0.48\linewidth}
\centering
\vspace{0pt}
\captionsetup{width=.95\linewidth}
\includegraphics[width=1.0\textwidth]{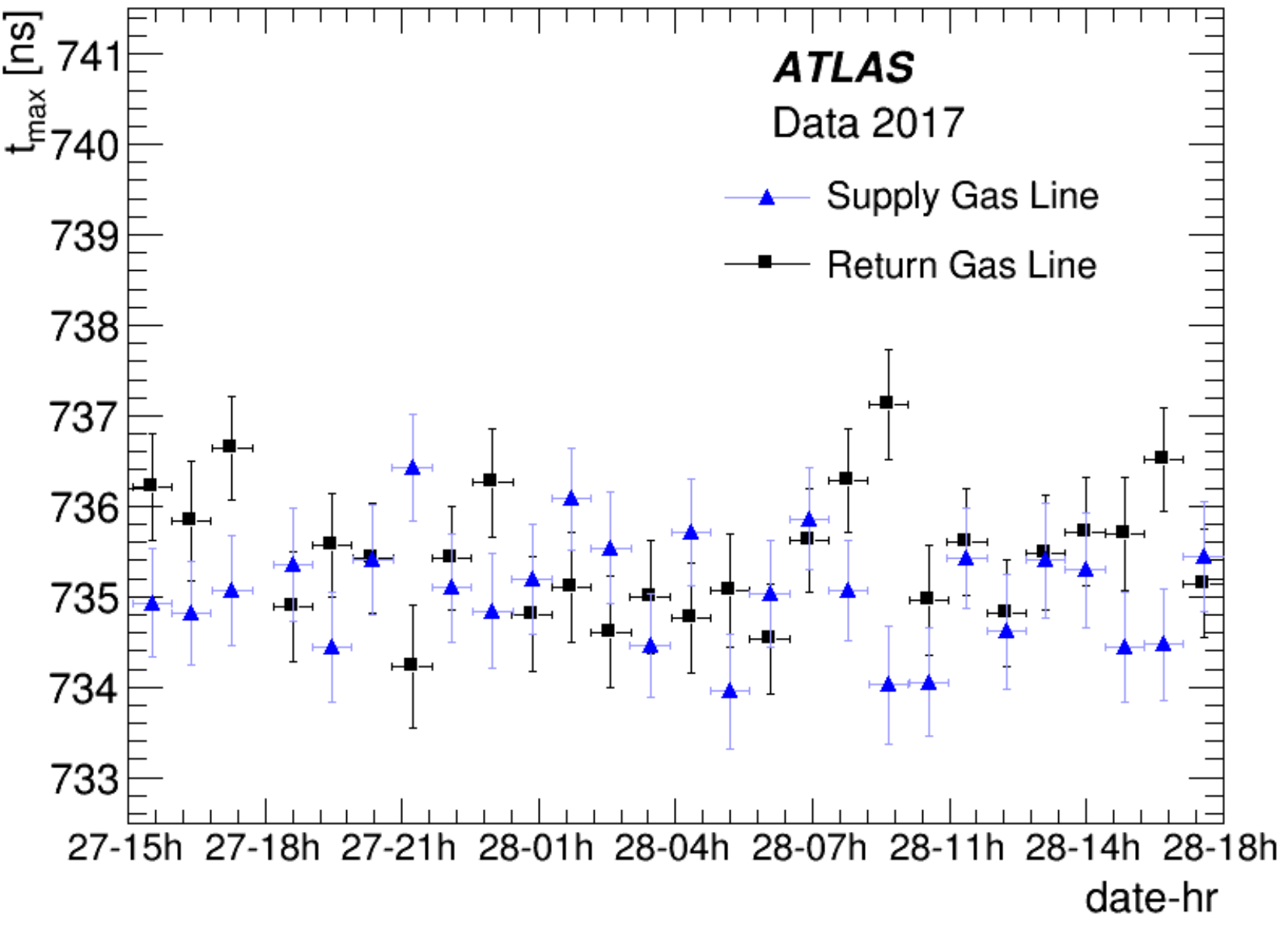}
\caption{Hourly GMC measurements of \TMAX~over a single day during
Run~2.}
\label{fig:TmaxDaily}
\end{minipage}
\hfill
\begin{minipage}[t]{0.48\linewidth}
\centering
\vspace{0pt}
\captionsetup{width=.95\linewidth}
\includegraphics[width=1.0\textwidth]{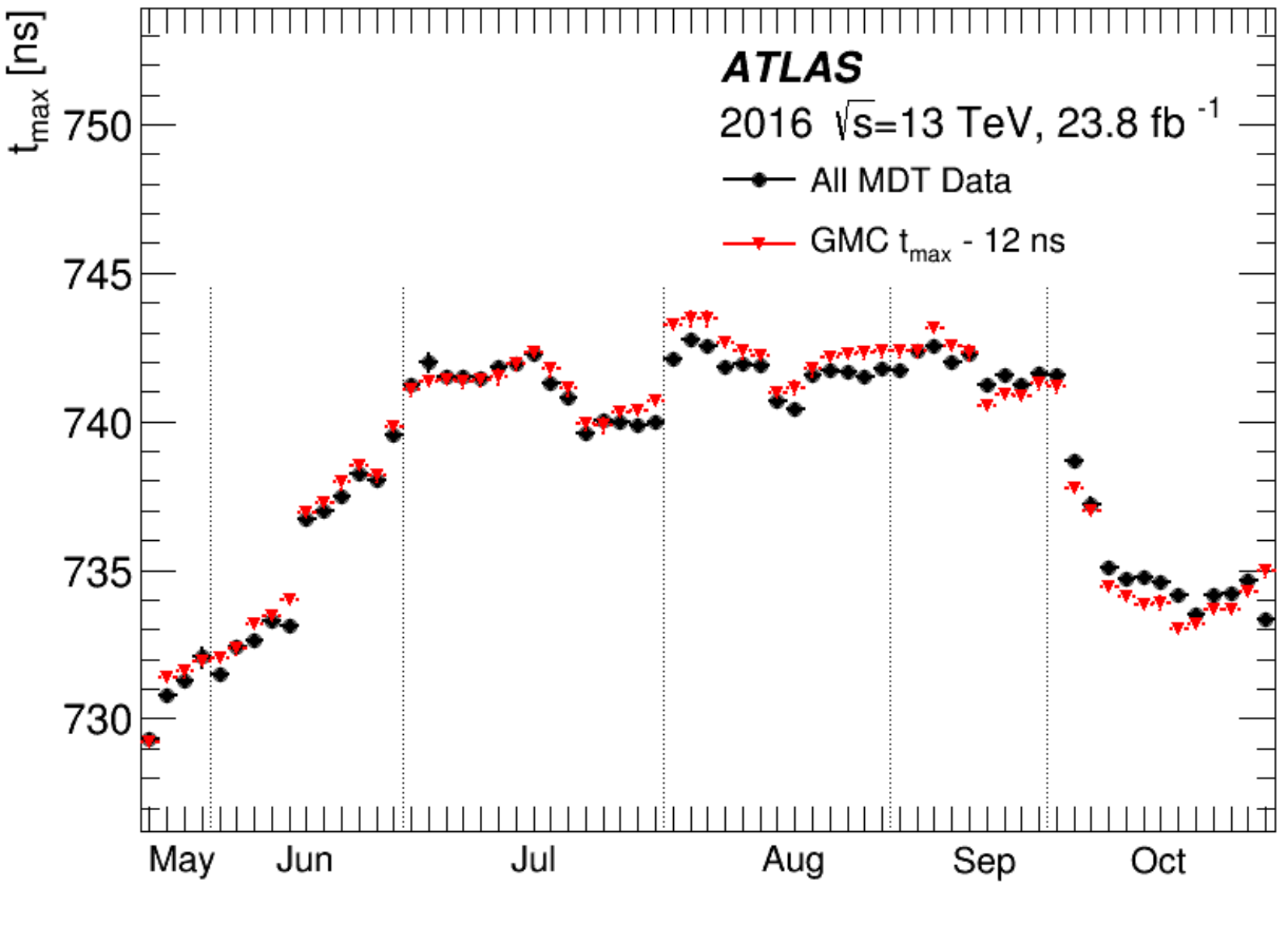}
\caption{Measurement of \TMAX~daily average over the entire 2016
data-taking period,
obtained from an average of all chambers and from the GMC.
The horizontal axis represents run numbers, which are not uniformly
distributed in time. Dotted lines delineate months.}
\label{fig:TmaxRun2-2016}
\end{minipage}
\end{figure}

The principal GMC output is a measurement of the maximum drift time,
\TMAX${}=t_\mathrm{w}^{}-$\TZ, where $t_\mathrm{w}^{}$ is the arrival time
of a hit located at the tube wall.
A drift time spectrum,  similar to Fig.~\ref{fig:TDCfit}, is computed for standard operating conditions by applying a
correction function~\cite{Amram:2011ue} based on the output of chamber-mounted
pressure and temperature sensors. The tail end of the spectrum is  fitted using a
modified Fermi-Dirac sigmoid function similar to that used to obtain the \TZ offsets, and the
\TMAX~is defined as a fit parameter, as described in Ref.~\cite{Levin:2008zza}.
The \TMAX~is published hourly to the online run monitoring system and to the muon calibration centers.
A variation of \TMAX$\mathrm{{} > 1}$~ns triggers the calculation of a new
set of \RT~functions for the track reconstruction.
The \TMAX~is sensitive to small changes in the gas composition: for example, a
100~ppm absolute increase in water vapour content yields a nearly 7~ns
increase in \TMAX~\cite{Amram:2008ola}.
A single 24-hour period of GMC \TMAX~data is shown in
Figure~\ref{fig:TmaxDaily}, where the stability is better than 1~ns RMS.
The GMC \TMAX~superimposed with the same quantity determined from the
average of all chambers over the 2016 data-taking period is presented in
Figure~\ref{fig:TmaxRun2-2016}. A one-time offset of $-$12~ns is applied to compensate
for the typical difference in gas temperature between the MDT and the standard
URT that range from about 4.5 to 5~K.
The range of \TMAX~variation over one year is 15~ns.
This change, observed annually, is attributed to the seasonal variation in
ambient humidity in the ATLAS cavern that modulates the content of water vapour
in the gas mixture.
 
\par
The GMC produces URT functions every two hours.
After local GMC temperature and pressure corrections are applied, the URT
registers only changes in the gas composition.
Based on the daily average of measurements from detector-mounted thermal
sensors, the URT is converted into chamber-specific
\RT~functions~\cite{Amram:2011ue}. These chamber \RT~functions are
further tuned by the $\delta$P correction parameter, as detailed in
Section~\ref{sec:dpcorrection}.
 
\subsection {Residuals and resolution} \label{sec:resolution}
\begin{figure}[htpb]
\begin{minipage}[t]{0.48\linewidth}
\centering
\vspace{0pt}
\captionsetup{width=.95\linewidth}
\includegraphics[width=1.0\textwidth]{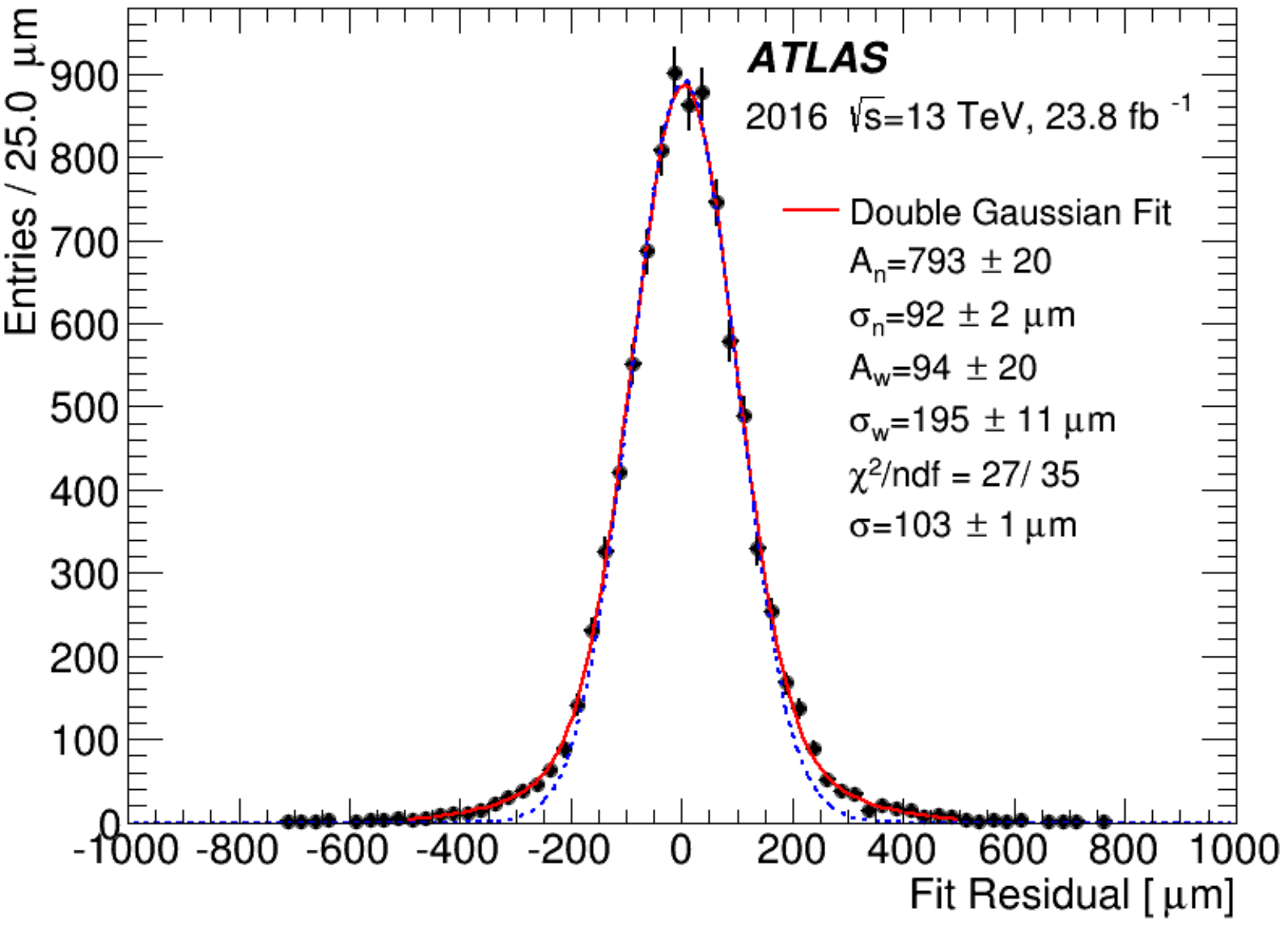}
\caption{Fit residual distribution for a representative chamber for
the radial bin $r=[2,3]$~mm. A double-Gaussian (solid line) is fit in the range
${} \pm$0.5~mm. A single-Gaussian (dotted line) is fit within the FWHM but shown
out to ${} \pm$0.5~mm.}
\label{fig:EML3C11_FitResi_mm0203}
\end{minipage}
\hfill
\begin{minipage}[t]{0.48\linewidth}
\centering
\vspace{0pt}
\captionsetup{width=.95\linewidth}
\includegraphics[width=1.0\textwidth]{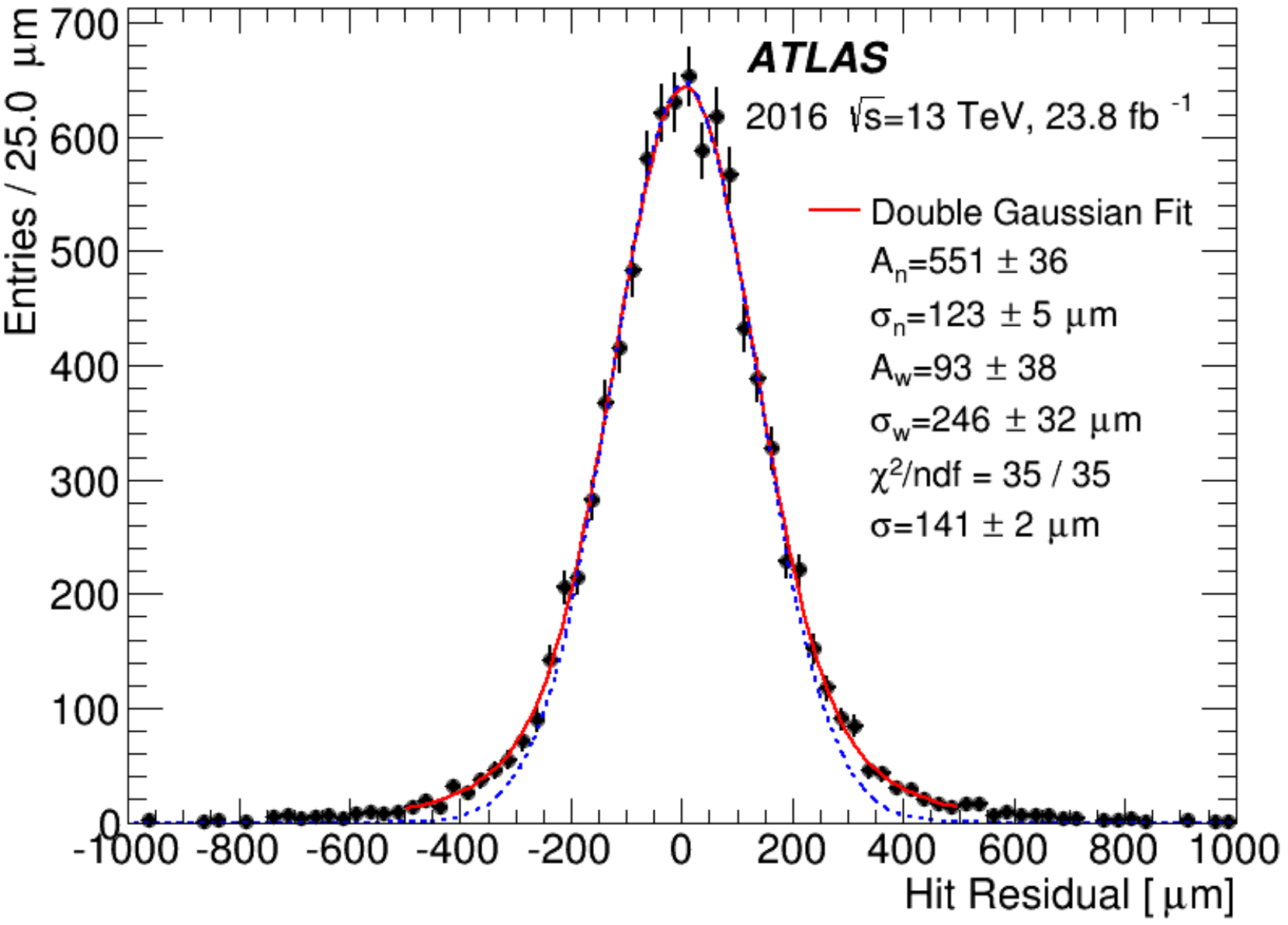}
\caption{Hit residual distribution for the same chamber and radial
bin as in Figure~\ref{fig:EML3C11_FitResi_mm0203}.}
\label{fig:EML3C11_HitResi_mm0203}
\end{minipage}
\end{figure}

Track segments in a chamber are the best-fit straight lines tangent to
the $N$ circles defined by $N$ drift radii.
The distance of  closest approach of a hit radius to this fit line
is either a \textit{fit} or \textit{hit} residual.
The former results when all $N$ data points are included in the linear fit,
and the latter is a measure of the $j^{\mathrm {th}}$ residual when the
$j^{\mathrm {th}}$ point is excluded from the fit of the other $N-1$ points.
Hit and fit residual distributions ($N$ entries per track, for all tracks in
the analysed data set) are generated for each MDT chamber.
Residuals are calculated for each of 15 radial bins, 1~mm wide, from 0 to 15~mm
in drift radius $ r $, as well as for a single 15-mm-wide radial bin.
The former are used to obtain the radial dependence of the resolution and the
latter is used to obtain the single-hit spatial resolution, both averaged over all
tubes in a chamber.
The residual distributions are fitted using a double-Gaussian function as
expressed in Equation~\ref{eq:doublegauss}:
This choice is dictated by the stochastic nature of the signal formation
that arises
from ion-pair creation, gas diffusion and electron--ion transport and,
importantly, allows for the dependence of the drift velocity on drift radius,
as indicated by the non-linearity of the \RT~function in
Figure~\ref{fig:rtfunction}.
These properties produce residual distributions with radially dependent
widths, including tails that are attributed to misassociated track hits
from delta-rays and asynchronous background hits.
While a single-Gaussian function inadequately describes the distributions for all
radial bins, the double-Gaussian function used here provides a good fit to the
residuals with a minimal increase in the number of free parameters.
The fit function is:
\begin{equation}
\label{eq:doublegauss}
f(r) = A_{\mathrm n}^{} \mathrm{e}^{\frac {-(r-\bar{r})^2} {2\sigma_{\mathrm n}^2} } +
A_{\mathrm w}^{} \mathrm{e}^{\frac {-(r-\bar{r})^2} {2\sigma_{\mathrm w}^2} } ,
\end{equation}
where $A_{\mathrm n}^{}$, $A_{\mathrm w}^{}$, $\sigma_{\mathrm n}^{}$,
$\sigma_{\mathrm w}^{}$ are the amplitudes and
standard deviations of a narrow and a wide Gaussian component.
The two terms in Equation~(\ref{eq:doublegauss}) are constrained to a common
mean $\bar r$, and fit over a residual range of ${}\pm 0.5$~mm, which contains
more than 99\% of hits (Section~\ref{sec:systematics}).
The $\sigma$ parameters (for fit and hit residuals) are determined for
the $ r^{\mathrm{th}}$ radial bin of the amplitude-weighted standard deviations
of the double-Gaussian function by:
\begin{equation}
\sigma(r) =
{A_{\mathrm n}^{}(r) \sigma_{\mathrm n}^{}(r) + A_{\mathrm w}^{}(r)
\sigma_{\mathrm w}^{}(r) \over A_{\mathrm n}^{}(r) + A_{\mathrm w}^{}(r) } .
\label{eq:wg}
\end{equation} where the dependence of the fit parameters on the drift radius is explicitly indicated.
Fit and hit residual distributions for a representative chamber at a drift
radius near the wire, where the resolution is broader than far away from the wire,
are shown in Figures~\ref{fig:EML3C11_FitResi_mm0203}
and~\ref{fig:EML3C11_HitResi_mm0203}, respectively.
These distributions are fit with the function of Equation~(\ref{eq:doublegauss}).
For comparison purposes a single-Gaussian fit within the FWHM of these distributions
is also shown.
 
The drift tube spatial resolution, $\mathrm{ \sigma_{res}^{} }$,
is the geometric mean of the fit and hit residual widths~\cite{Carnegie:2006hf}
defined as:
\begin{equation}
\sigma_{\mathrm{res}}(r) = \sqrt{ \sigma_{\mathrm{fit}}(r) \cdot \sigma_{\mathrm{hit}}(r) }
\label{eq:res}
\end{equation}
This approach was checked by Monte Carlo simulation.
About $\mathrm 10^6$ tracks incident on a generic chamber within a
$\mathrm 10^\circ$ angular range were generated.
The drift radii assigned to these tracks were Gaussian distributed
about their initial values using a radially dependent input resolution
function derived from a test beam~\cite{Deile:2016ddq}.
These smeared hit points were linearly fit and the output resolution
for each radial bin was obtained using Equation~(\ref{eq:res}).
The mean difference over the tube radius between the output and input
resolution is $\mathrm 0.3 \pm 0.2 \; \upmu$m, confirming the validity
of the formulation.

\section{Correction algorithms}
 
This section describes the algorithms used to determine the \DTZ~and
\DP~corrections and their associated metrics, the track residuals and the
resolution.
 
\subsection {\DTZ~correction}
\label{sec:T0correction}
 
\begin{figure}[h]
\begin{minipage}[t]{0.48\linewidth}
\centering
\vspace{0pt}
\captionsetup{width=.95\linewidth}
\includegraphics[width=1.0\linewidth]{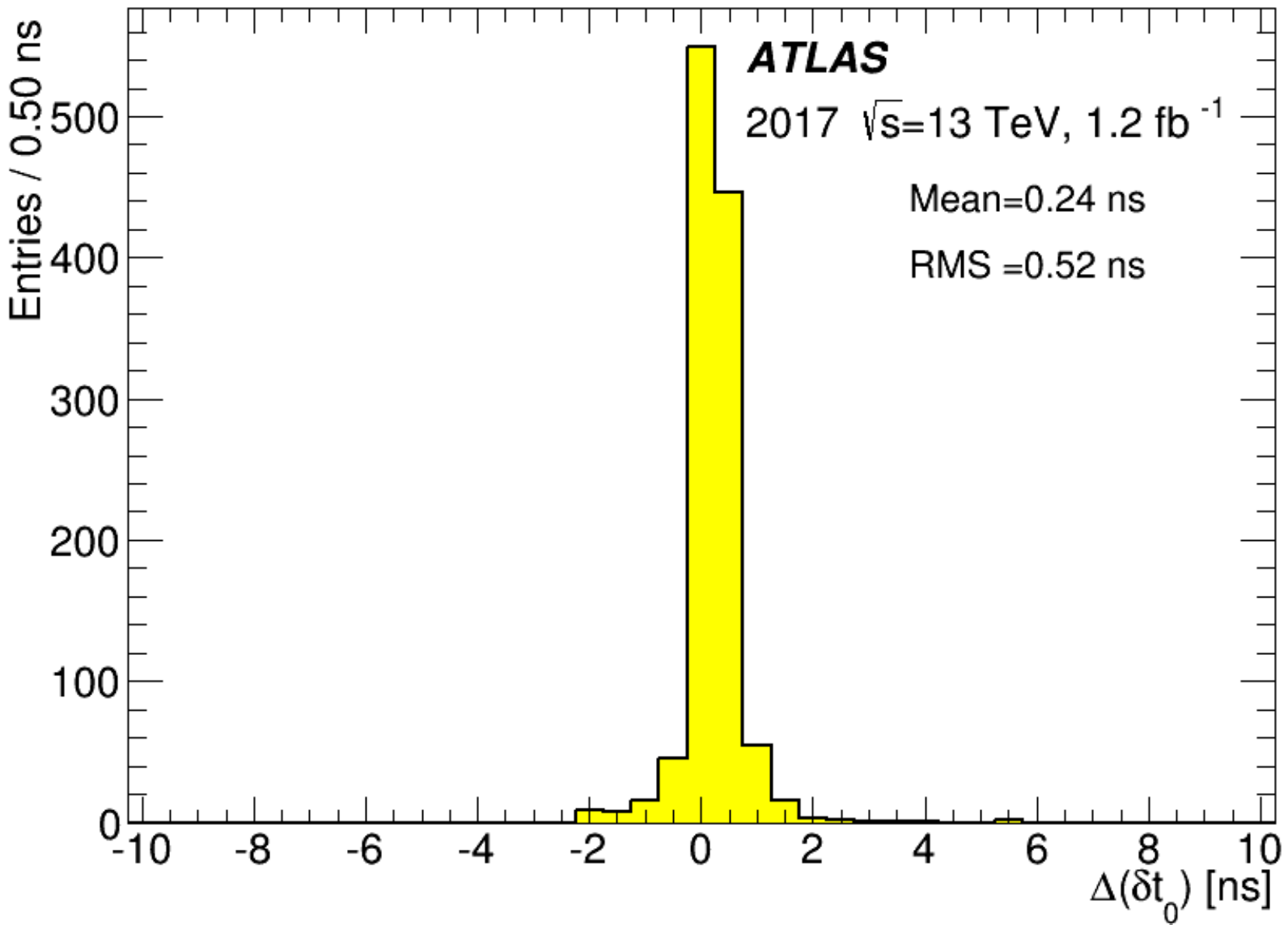} 
\caption{Chamber-by-chamber difference in the \DTZ~parameter obtained
from the June (toroid magnetic field off) and
November (toroid magnetic field on) 2017 data.}
\label{fig:DT_DIFFN_2017}
\end{minipage}
\hfill
\begin{minipage}[t]{0.48\linewidth}
\centering
\vspace{0pt}
\captionsetup{width=.95\linewidth}
\includegraphics[width=1.0\linewidth]{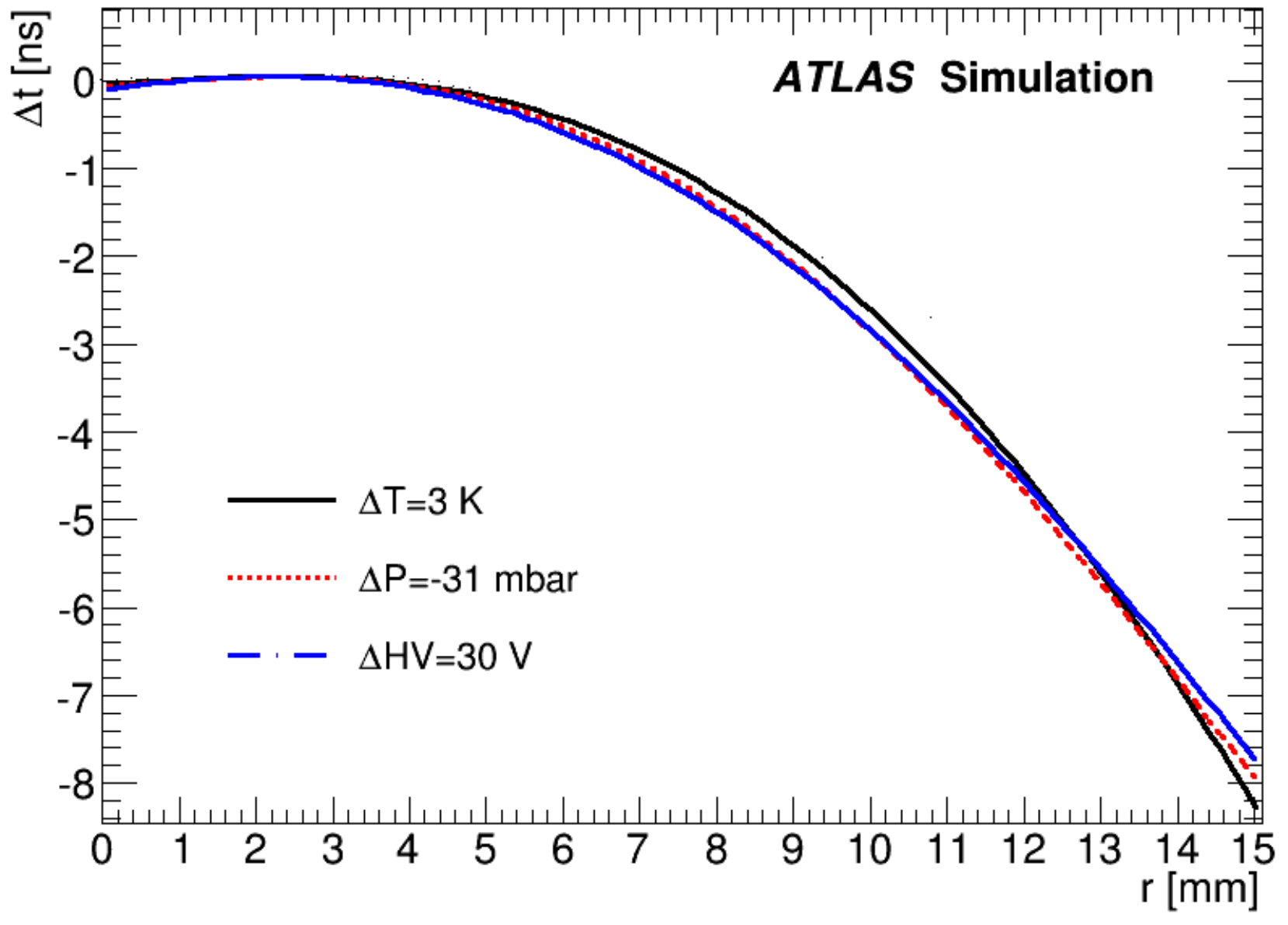}
\caption{Garfield simulation of drift time variation as a function of
drift radius for:
$+$3~K, $-$31~mbar, and $+$30~V change relative to
293~K, 3~bar and 3080~V,
respectively.}
\label{fig:rtTPTest}
\end{minipage}
\end{figure}

The default \TZ~offset definition may vary by the order of 1~ns from the \TZ~value that  provides the best resolution.
The algorithm described here is a variant of the tuning
procedures designed to find the best linear fit to a
track segment by tuning the \TZ~on a track-by-track basis~\cite{Branchini}.
The \DTZ~correction method instead optimizes the default \TZ~by testing a
new, modified \TZ~associated with a single chamber simultaneously for all
tubes and for all tracks in a data run, while keeping the \RT~function fixed.
The algorithm operates on track segments containing at least as many hits
as there are drift tube layers in the chamber, i.e.
a minimum of six (eight) hits for a six-layer (eight-layer) chamber.
The procedure steps through \DTZ~over a range of $\pm 10$~ns in 0.5~ns steps.
At each step, new drift times and drift radii are computed and the segments
are re-fit.
Importantly, the residuals are evaluated only for hits with drift radii
$r<5$~mm  where the drift velocity is largest, about $\mathrm 50\; \upmu$m/ns.
By using only these near-wire residuals the procedure is practically insensitive
to the \RT~function and avoids compensating for systematic uncertainties which
come into play at larger drift radii (see Figure~\ref{fig:rtTPTest}).
 
\begin{figure}[htpb]
\begin{minipage}[t]{0.48\linewidth}
\centering
\vspace{0pt}
\captionsetup{width=.95\linewidth}
\includegraphics[width=1.0\textwidth]{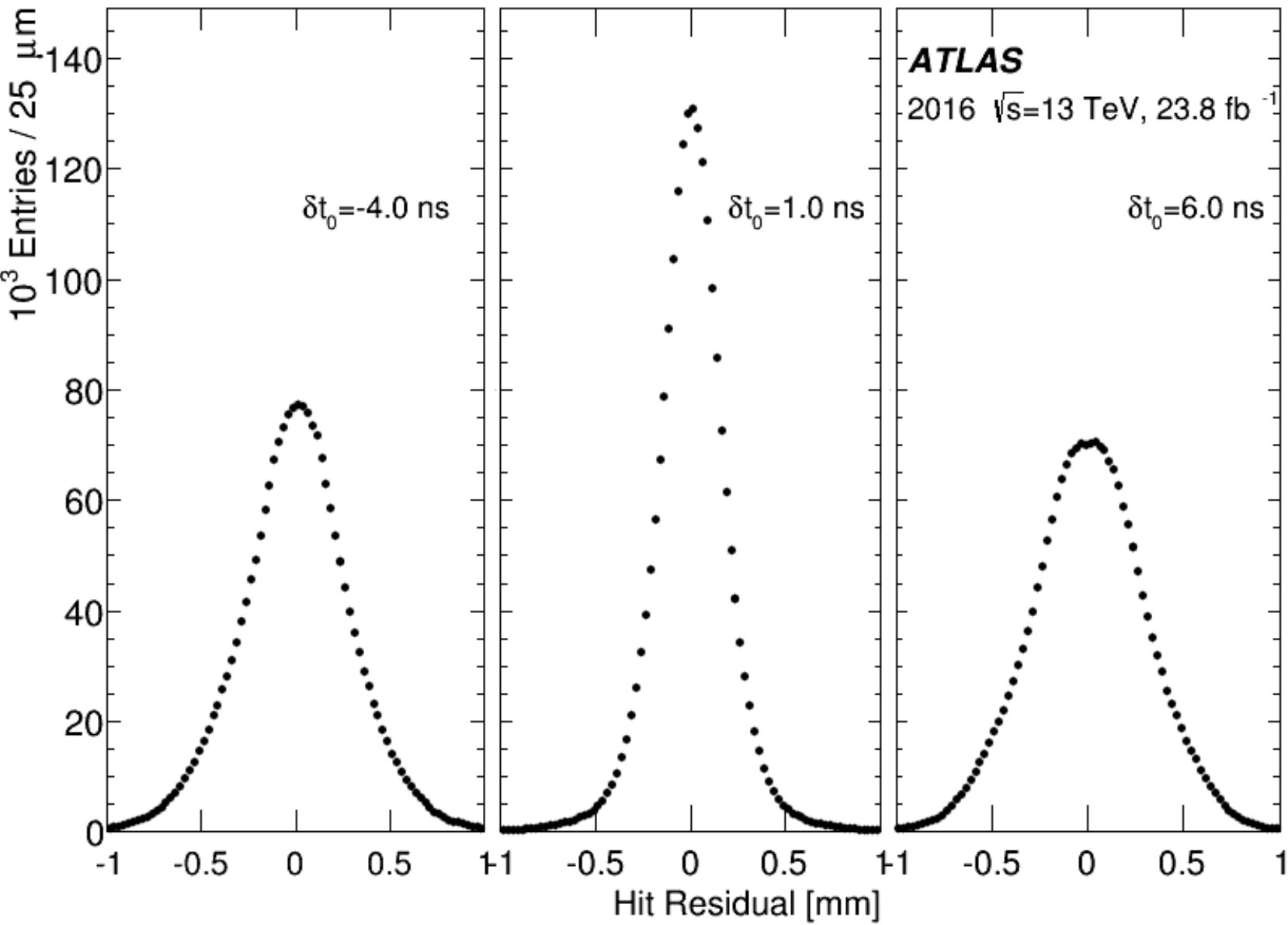}
\caption{Hit residual distributions for a representative chamber for
\TZ~correction parameter set to $-$4~ns (left), $+1$~ns (center)
and $+$6~ns (right) from the default value.
Widths are 296, 188 and 303~$\upmu$m, respectively.}
\label{fig:EML3C11_HitResi_3DT}
\end{minipage}
\hfill
\begin{minipage}[t]{0.48\linewidth}
\centering
\vspace{0pt}
\captionsetup{width=.95\linewidth}
\includegraphics[width=1.0\textwidth]{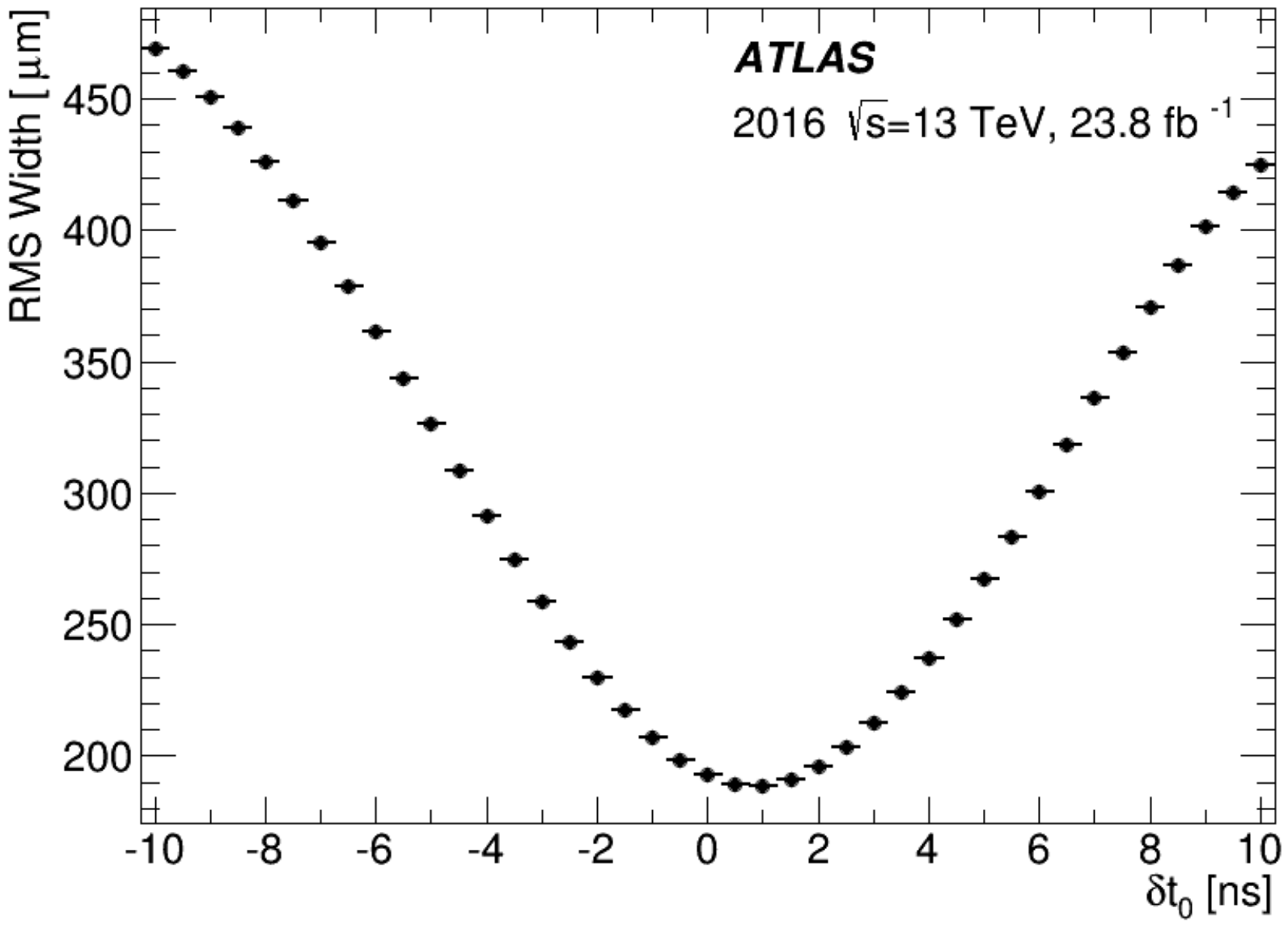}
\caption{Typical RMS widths of hit residual distributions of
hits less than 5~mm from the wire as a function of the \TZ~correction,
for the same chamber as in Figure~\ref{fig:EML3C11_HitResi_3DT}.}
\label{fig:EML3C11_HitResiVsDT}
\end{minipage}
\end{figure}
 
Figure~\ref{fig:EML3C11_HitResi_3DT} shows segment hit residuals from the
\TZ~correction procedure for a representative chamber for \DTZ~values
of $-$4, $+$1, and $+$6~ns.
Figure~\ref{fig:EML3C11_HitResiVsDT} shows the incremental change with each
step of the residual RMS widths for the entire range of \DTZ~values relative
to the initial \TZ.
The optimal \DTZ~is the one yielding the minima of the RMS widths of the
fit and hit residual distributions.
The value on the vertical axis corresponding to the minimum in
Figure~\ref{fig:EML3C11_HitResiVsDT} does not represent the resolution,
but only the RMS spread of the hit residuals near the wire.
\footnote{
For over 95\% of the MDT chambers the double-Gaussian fits yield the
same \TZ~minima as obtained using the RMS of the residual distributions.
In the remaining chambers the fit method can lead to anomalous minima, so
for consistency and stability of the results, the RMS method to establish
the minimum is adopted for all chambers.
}
\par
\begin{figure}[h]
\begin{minipage}[t]{0.48\linewidth}
\centering
\vspace{0pt}
\captionsetup{width=.95\linewidth}
\includegraphics[width=1.0\textwidth]{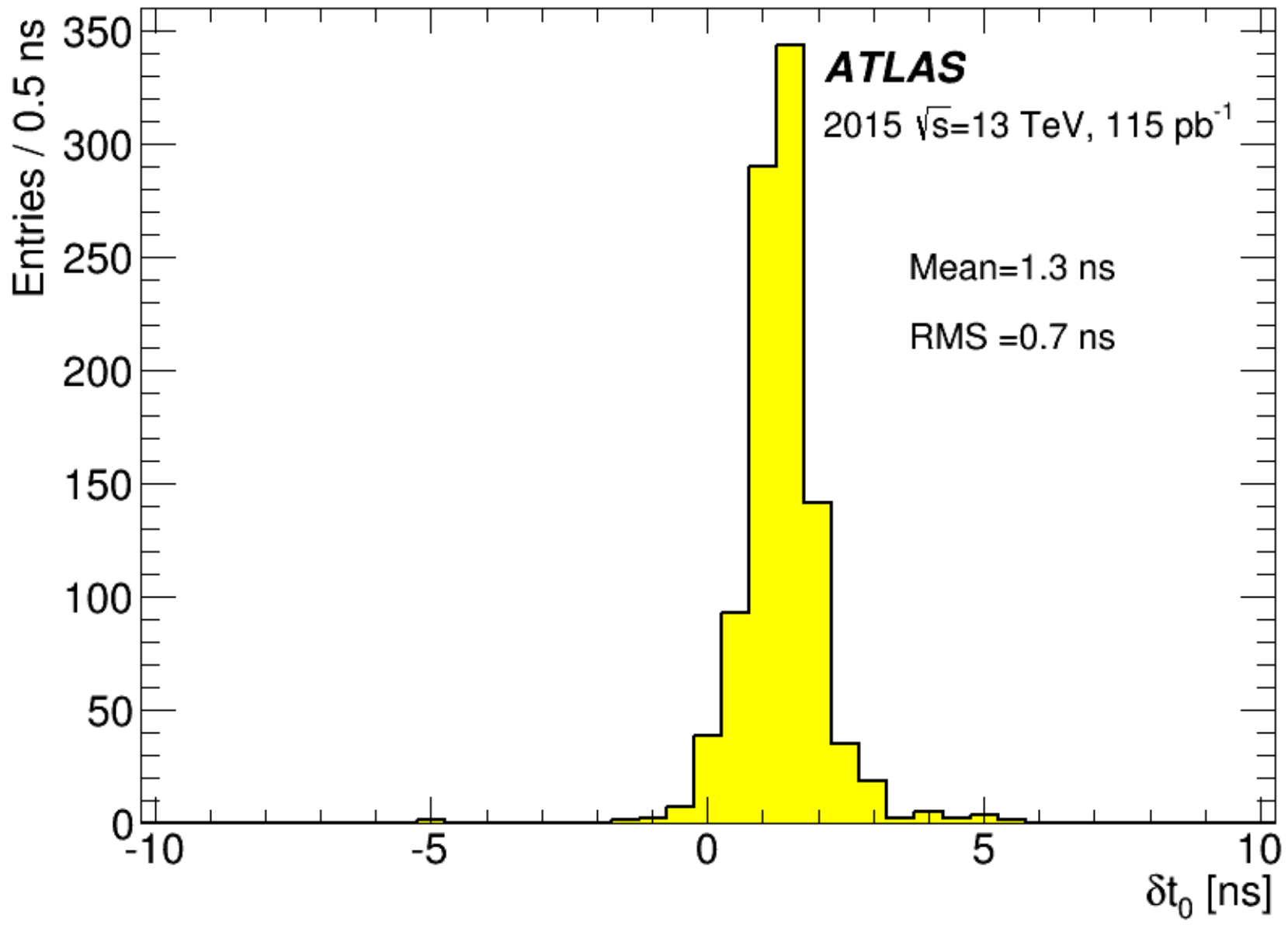}
\caption{Distribution of \DTZ~for one run in November 2015
determined by the \TZ~correction algorithm.}
\label{fig:deltaT0November}
\end{minipage}
\hfill
\begin{minipage}[t]{0.48\linewidth}
\centering
\vspace{0pt}
\captionsetup{width=.95\linewidth}
\includegraphics[width=1.0\textwidth]{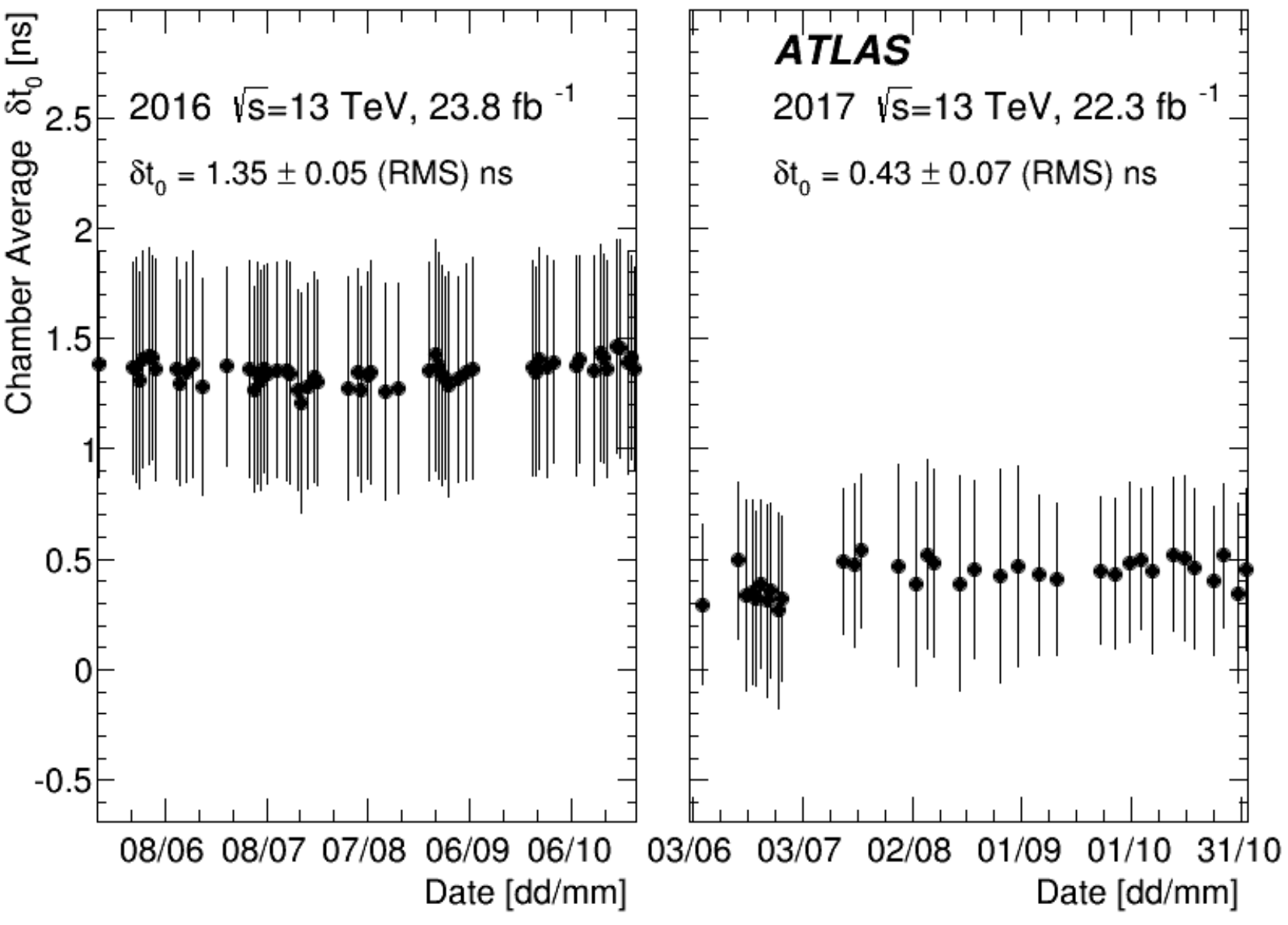}
\caption{Chamber-averaged \DTZ~values for runs in 2016 (left) and 2017 (right).
Uncertainties are the RMS of the distributions.
}
\label{fig:DT_AVG_TIME_2016}
\end{minipage}
\end{figure}

The \DTZ~correction procedure is applied to 2015 data (the first year of
Run~2 operation) using initial \TZ~offsets generated by the MDT
calibration centers.
The \DTZ~corrections obtained in a much higher peak-luminosity November 2015 run,
$\mathrm {4.8 \times 10^{33} \; \mathrm{cm^{-2}} \; \mathrm{s^{-1}} }$, are shown in
Figure~\ref{fig:deltaT0November}, which reveals an average \TZ~global
correction of 1.3~ns with a RMS spread of 0.7~ns.
The tails of the distribution are populated by a few chambers that were
measured to have been constructed with structural non-conformities affecting
the internal alignment of the tubes and therefore of the hit positions.
The \DTZ~correction tries to compensate for this misalignment.
 
\par
The determination of the \DTZ~corrections for all MDT chambers was repeated for
a selection of data runs that span all data-taking periods in 2016 and 2017.
The chamber-averaged \DTZ~corrections for each run are shown in
Figure~\ref{fig:DT_AVG_TIME_2016}.
At least 1165 out of the total 1172 chambers are included in each data point
used for this average, depending on the integrated luminosity of the run.
For the 65 runs analysed in  2016 the mean global \DTZ~offset is
$\mathrm{1.35 \pm 0.05 }$~(RMS)~ns,
consistent with the result determined from the 2015 data set.
The \DTZ~corrections from 2016 are then added to the calibration constants for 2017,
so that the algorithm, when applied to the 2017 data set, produces a
smaller global shift with a mean of $\mathrm{ 0.43 \pm 0.07 }$~(RMS)~ns.

Figure~\ref{fig:DT_DIFFN_2017} shows the chamber-by-chamber \DTZ~difference
between the first and last runs of the 2017 data set.
In the first run, the toroid magnetic field was off and
in the last run the field was at full strength and the instantaneous
luminosity was higher by one order of magnitude.
The mean difference is consistent with zero.
This result demonstrates that the \DTZ~parameter, once established,
remains stable for all chambers over a large range of instantaneous
luminosity and is insensitive to the magnetic field.

\subsection {\DP~correction}
\label{sec:dpcorrection}
Optimization of the \RT~functions is done with an \textit {effective pressure}
parameter, \DP, that corrects for the (small) inaccuracies in
gas pressure sensors, temperature measurement, and high voltage.
In all three cases, minor variations from nominal values produce  
radially dependent changes in the \RT~function, such that a single, scalable correction
function can be used.
The calculation of this function is done using the
Garfield program~\cite{GARF} configured to model the drift electron
transport in MDT tubes with standard operating parameters as in
Section~\ref{MSdescription}.
Specifically, a pure Ar--CO$_2$ gas mixture without water vapour and
zero magnetic field is used. Figure~\ref{fig:rtTPTest} shows the predicted
change in the drift time as a function of drift radius for variations of
temperature, pressure and high voltage from their reference values.
This drift time response  was previously validated
with temperature data~\cite{Amram:2011ue}.
The maximum effect on the drift time is at the tube wall as expressed in
Equation~(\ref{eq:DtAtWall}),
where the temperature correction is $-$2.6~ns/K, while the
pressure correction function scales as
$ -0.1$~K/mbar relative to the temperature correction.
The negative sign means that an increase in gas pressure at fixed temperature
corresponds to an increase in gas density, a decreased drift electron mean
free path, and thus longer drift times.
Similarly, a higher voltage at fixed pressure leads to an increased electric
field  and higher drift velocity.
Importantly, for any of these parameters, the curves in
Figure~\ref{fig:rtTPTest} indicate a negligible effect on drift times
for radii smaller than 5~mm.
For small deviations of the high voltage from the nominal setting, the drift
time changes by only about 0.26~ns/V at the tube wall, thus typical systematic
offsets of 1 V have a negligible effect:
\begin{equation}
\label{eq:DtAtWall}
\Delta t\mathrm{(wall)~[ns]} = -2.6  \, \frac{\mathrm{ns}}{\mathrm{K}} \times \Delta T~\mathrm{[K]}
= +0.26 \, \frac{\mathrm{ns}}{\mathrm{mbar}} \times
\Delta P~\mathrm{[mbar]}
= -0.26 \, \frac{\mathrm{ns}}{\mathrm{V}} \times
\Delta V~\mathrm{[V]} .
\end{equation}

\par
The effective pressure correction algorithm is similar to the \DTZ~method
and uses the same track-segment hit selection.
For a collection of track-segment hits, \DP~is incremented in $5$~mbar
steps through a ${} \pm$100~mbar range of test pressures centered around
the nominal 3~bar.
The \RT~function is recalculated using the correction function shown
in Figure~\ref{fig:rtTPTest} rescaled by the $\mathrm P + \delta$P
effective pressure value assumed in the scan, new drift radii are obtained
from the newly corrected \RT~function, and track segments are re-fit and
residuals evaluated.
The procedure is initiated with the set of chamber-specific \RT~functions
of the day of the run.
The \DTZ~corrections found previously are applied to the hit drift times.
For \DP~determination, the residuals are calculated for all drift radii.
 
\par
\begin{figure}[htpb]
\begin{minipage}[t]{0.48\linewidth}
\centering
\vspace{0pt}
\captionsetup{width=.95\linewidth}
\includegraphics[width=1.0\textwidth]{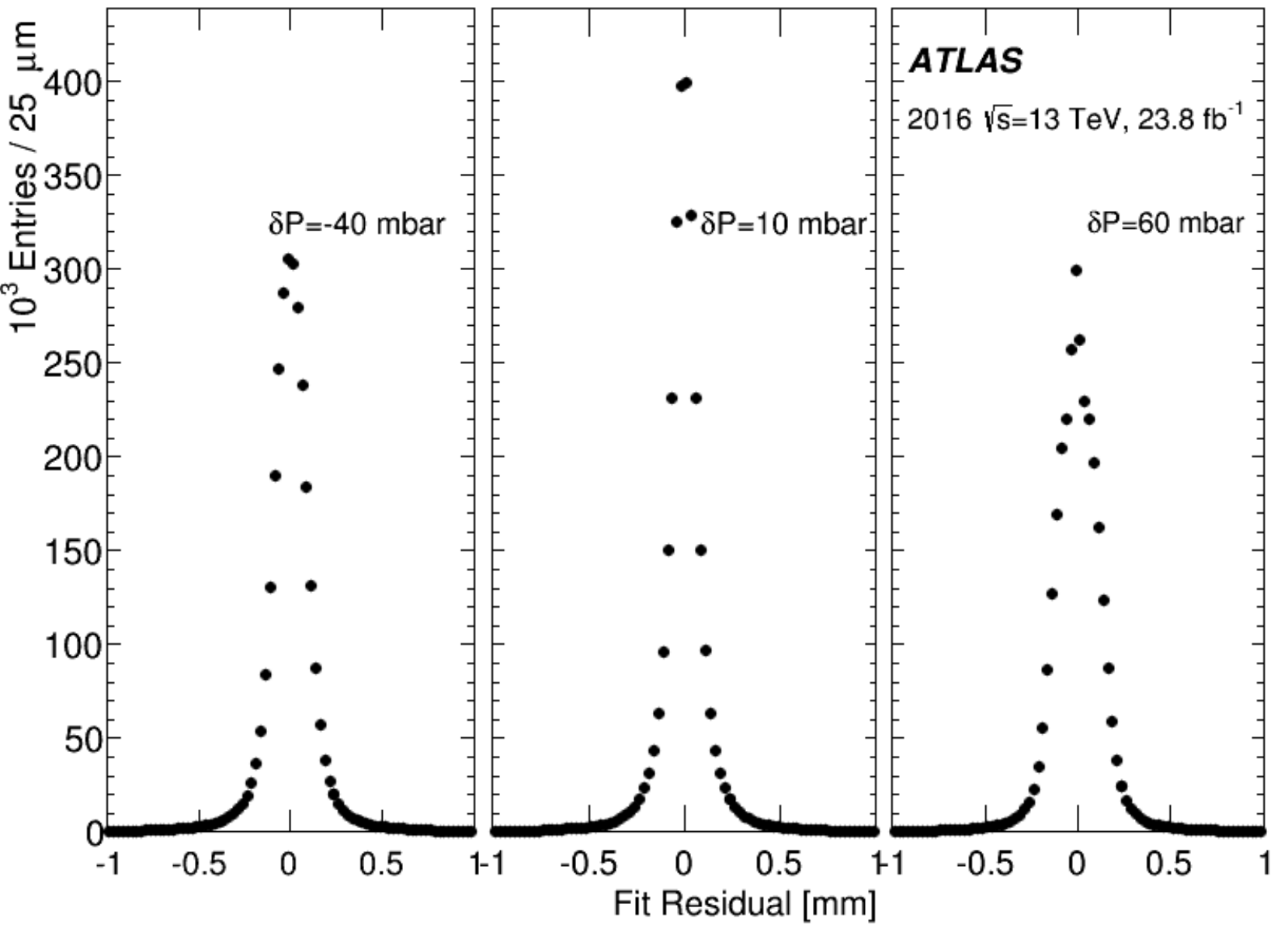}
\caption{Fit residual distributions for a representative MDT
chamber for \DP$\mathrm{{}=-40~mbar}$ (left),
$+$10~mbar (middle, optimal setting) and $+$60~mbar (right).
RMS widths are 141, 131 and 144~$\upmu$m, respectively}
\label{fig:HitResi_3DP}
\end{minipage}
\hfill
\begin{minipage}[t]{0.48\linewidth}
\centering
\vspace{0pt}
\captionsetup{width=.95\linewidth}
\includegraphics[width=1.0\textwidth]{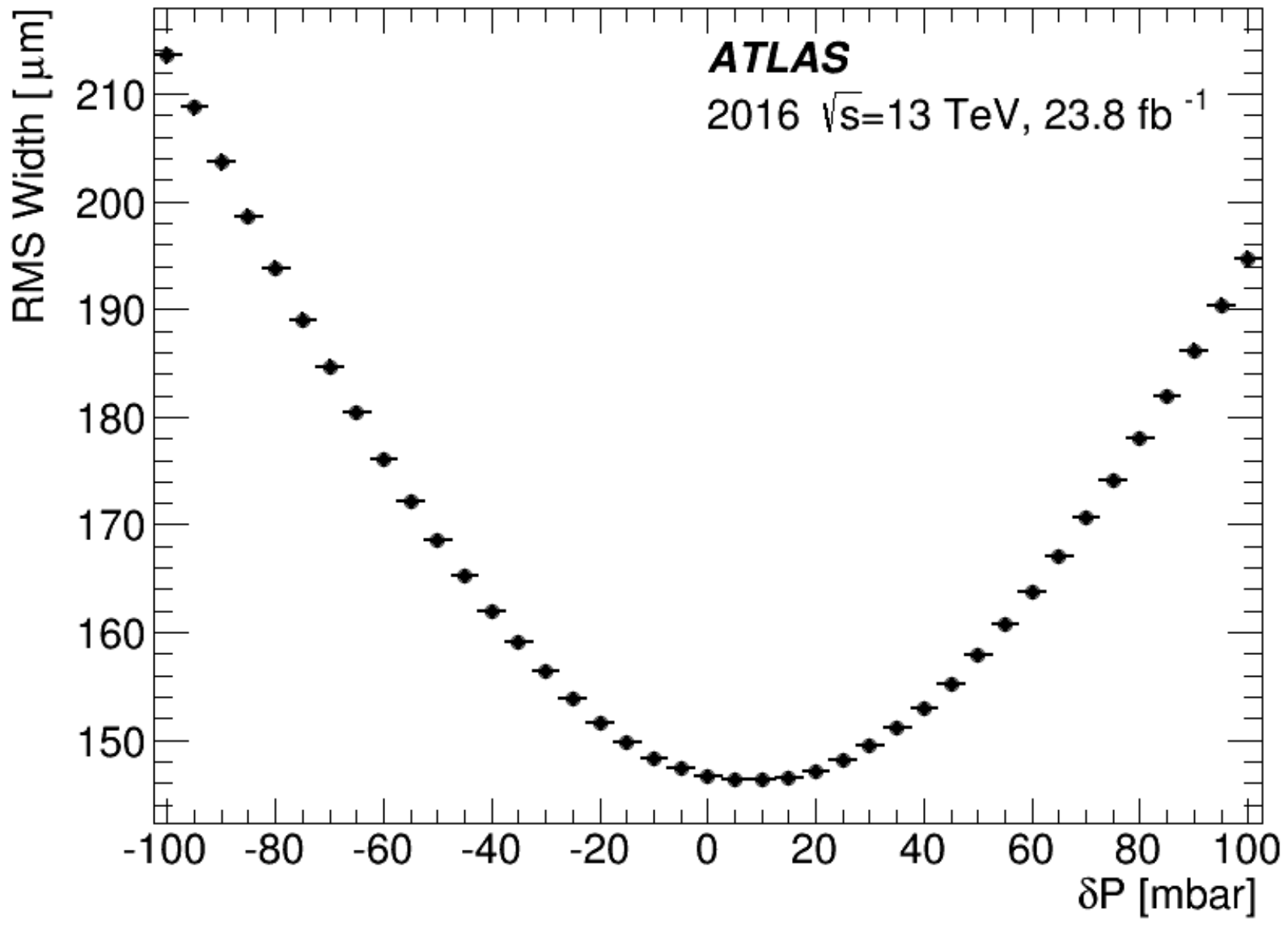}
\caption{Fit residual RMS as a function of \DP~for the same chamber as in
Figure~\ref{fig:HitResi_3DP}.}
\label{fig:hitresvsdP}
\end{minipage}
\end{figure}

Similar to the \DTZ~procedure, the optimal \DP~is the one yielding the
minimum of the RMS width of the fit and hit residual distributions.
The hit residuals for three \DP~values, of $-$40~mbar, 10~mbar and 60~mbar,
are shown for a representative chamber in Figure~\ref{fig:HitResi_3DP}.
Figure~\ref{fig:hitresvsdP} shows the RMS width of the hit residual
distribution as a function of \DP~for the full scan
range.
The minimum at 10~mbar represents the best value of this parameter.
 
\par
\begin{figure}[htpb]
\begin{minipage}[t]{0.48\linewidth}
\centering
\vspace{0pt}
\captionsetup{width=.95\linewidth}
\includegraphics[width=1.0\textwidth]{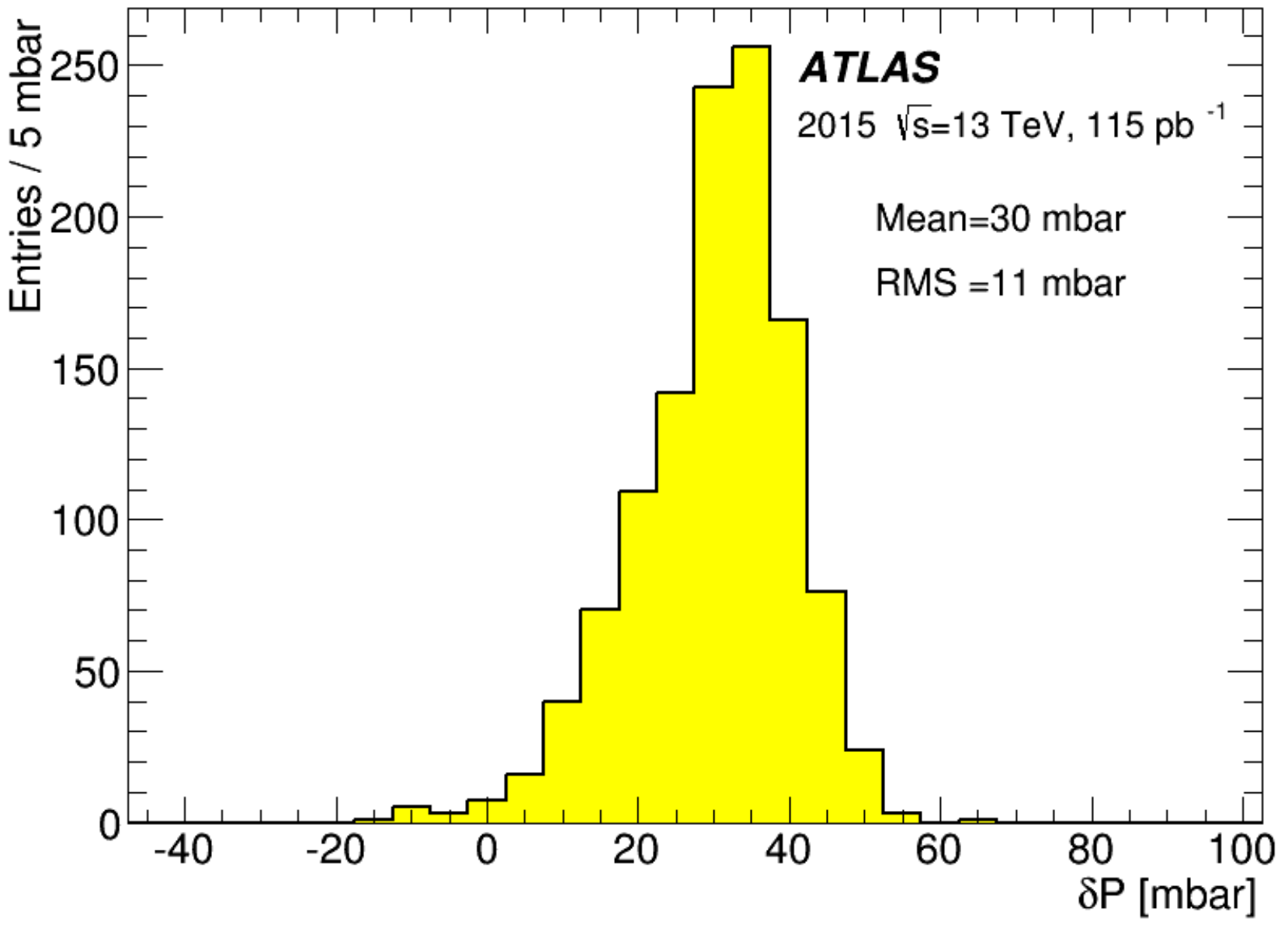}
\caption{Distribution of \DP~for a high-luminosity run in November 2015.}
\label{fig:dPToroidOnOff}
\end{minipage}
\hfill
\begin{minipage}[t]{0.48\linewidth}
\centering
\vspace{0pt}
\captionsetup{width=.95\linewidth}
\includegraphics[width=1.0\textwidth]{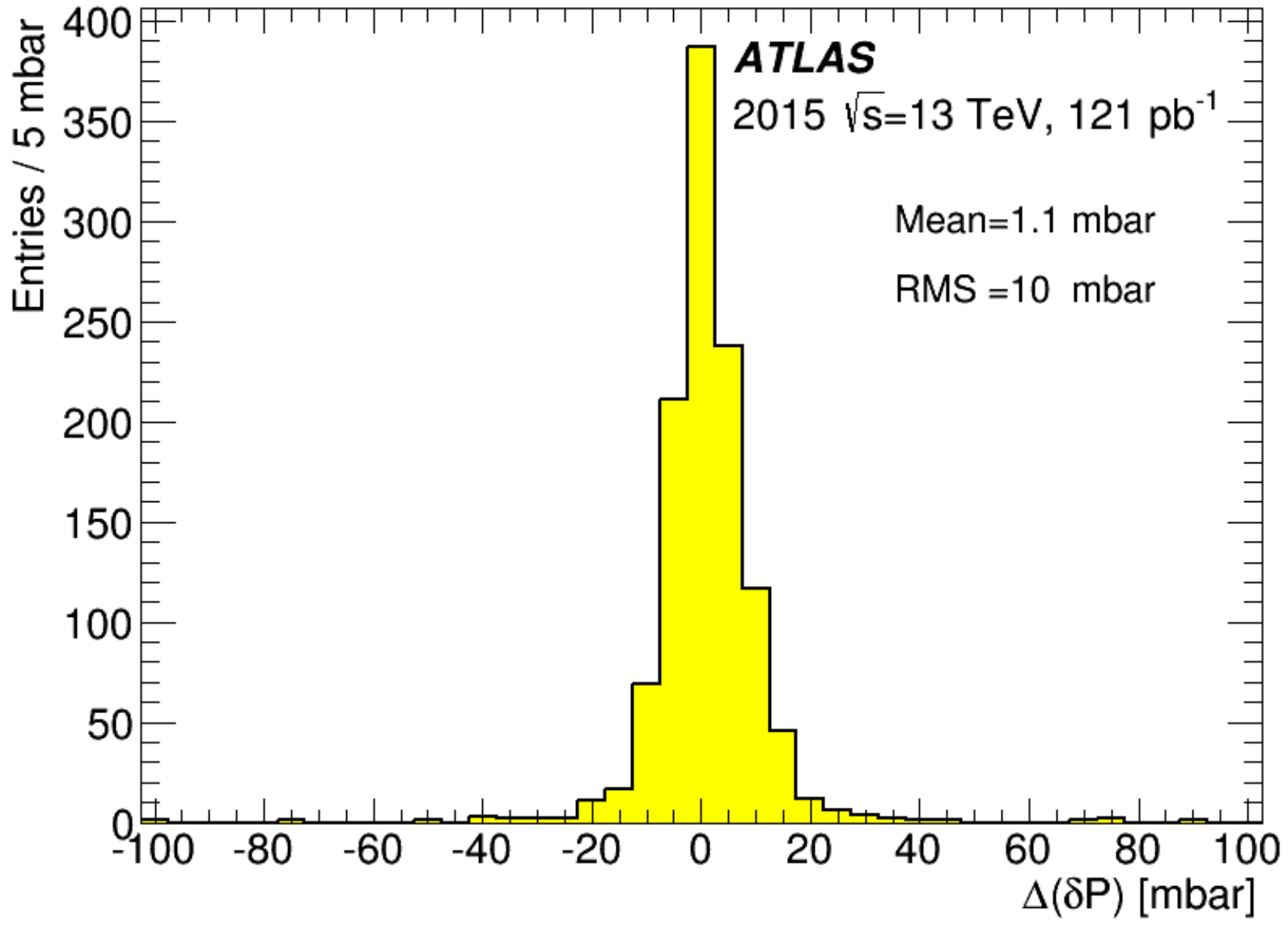}
\caption{Chamber-by-chamber difference of \DP~for a run with the
toroidal magnetic field off and a run 2.5 months later with 25
times higher luminosity and the toroid on.}
\label{fig:dPdiff}
\end{minipage}
\end{figure}

The minima of curves similar to Figure~\ref{fig:hitresvsdP}, taken as the
\DP~step corresponding to the lowest residual, were initially
determined for all MDT chambers using data from August 2015 with the toroid
magnet turned off.
The algorithm was applied to a set of collision runs acquired 2.5 months
later with the magnetic field turned on, and with 25 times higher
instantaneous luminosity (from $\mathrm{ 0.17 \times 10^{33} \ cm^{-2} \; s^{-1} }$
in mid August to $\mathrm{ 4.8 \times 10^{33} \ cm^{-2} \; s^{-1} }$ at beginning
of November), yielding in both cases an average correction \DP\ of
$\mathrm{ 30.0 \pm 0.3 }$~mbar with an 11~mbar RMS spread.
Distributions of the latter \DP~results are shown in
Figure~\ref{fig:dPToroidOnOff}, and the chamber-by-chamber differences
in the \DP~correction for the two data sets are shown in
Figure~\ref{fig:dPdiff}.
The measured change is within the \DP~step size.
The corrections are stable in time and insensitive to both the magnetic field
and the increased hit rates that result from higher instantaneous luminosity.
 
\begin{figure}[h]
\begin{minipage}[t]{0.48\linewidth}
\centering
\vspace{0pt}
\captionsetup{width=.95\linewidth}
\includegraphics[width=1.0\textwidth]{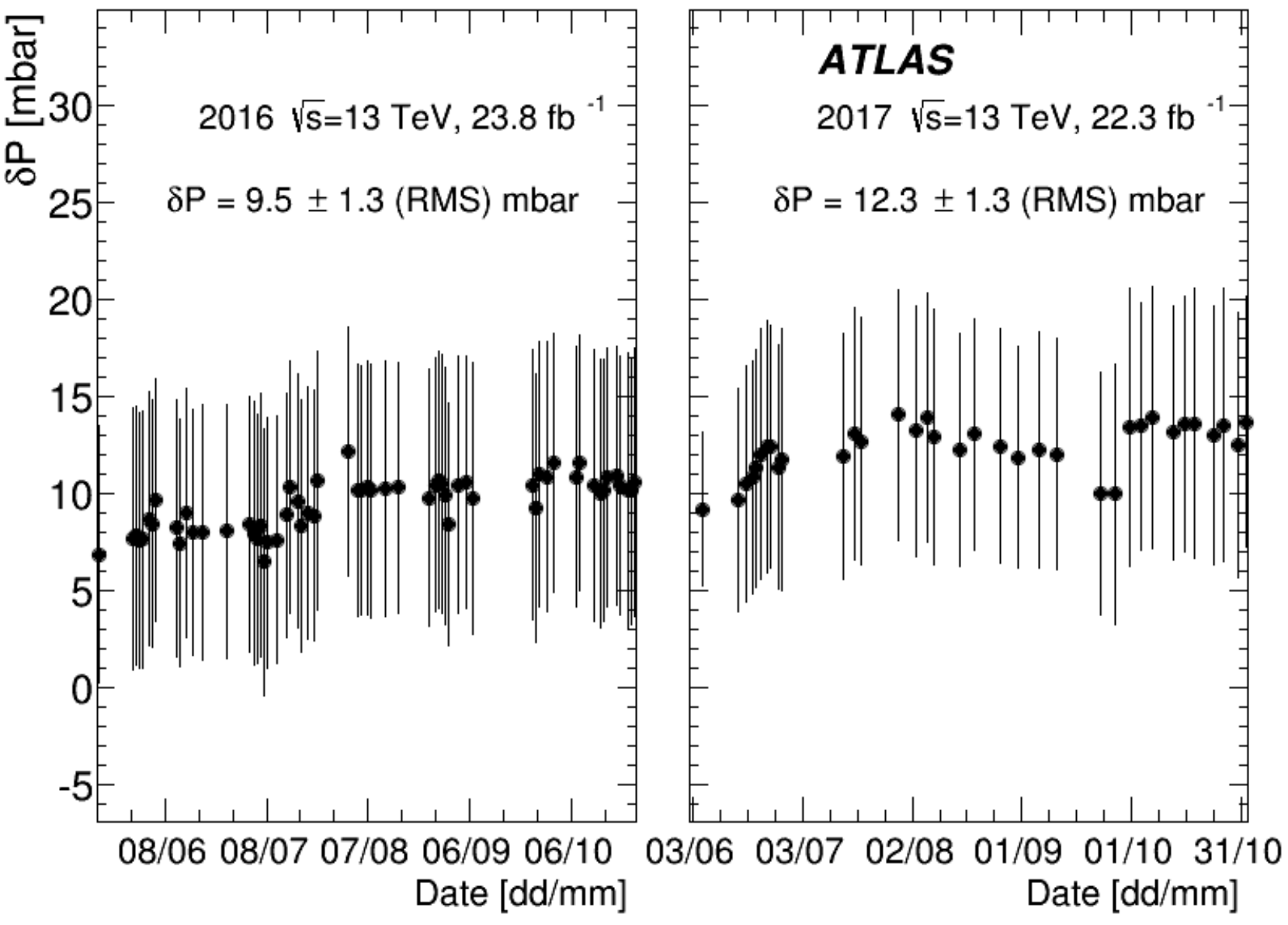}
\caption{Chamber-averaged \DP~for each analysed run in 2016
(left) and 2017 (right). The uncertainties are the RMS spread of the
chamber's \DP~distribution.  }
\label{fig:DP_AVG_TIME_2016}
\end{minipage}
\hfill
\begin{minipage}[t]{0.48\linewidth}
\centering
\vspace{0pt}
\captionsetup{width=.95\linewidth}
\includegraphics[width=1.0\textwidth]{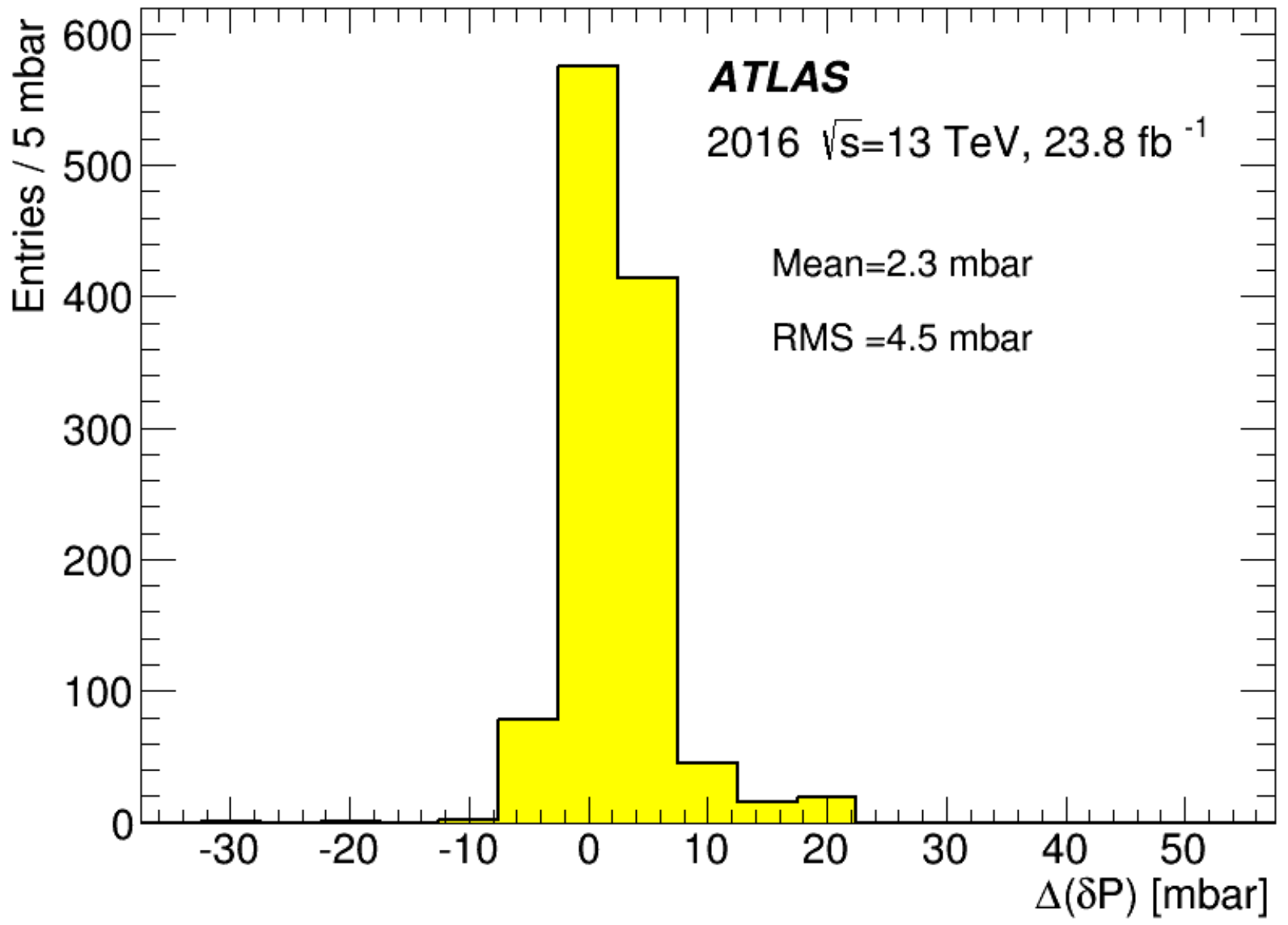}
\caption{Chamber-by-chamber differences in \DP~between the
first and last 2016 analysed run.}
\label{fig:DeltaDeltaP2016}
\end{minipage}
\end{figure}

The \DP~algorithm was further applied to 65 collision runs acquired from
May to October 2016, where the peak luminosity exceeded
$10^{34} \ \mathrm{cm}^{-2} \; \mathrm{s}^{-1}$,
more than twice that of November 2015.
These runs used \RT~functions which included the \DP~corrections determined
in the 2015 data.
The chamber-averaged \DP~value for each of the runs in 2016 and 2017 is shown
in Figure~\ref{fig:DP_AVG_TIME_2016} as a function of run date.
The average of \DP~over all of 2016 data is
$\mathrm{ 9.5 \pm 0.8 }$~mbar (RMS = 1.3~mbar) and
for 2017 is $\mathrm{ 12.3 \pm 1.0 }$~mbar
(RMS = 1.3~mbar).
The \DP~correction is stable over each year to better than 3~mbar.
Figure~\ref{fig:DeltaDeltaP2016} shows the distribution of the
chamber-by-chamber \DP~difference between the first and last runs
in the 2016 data set.
The mean difference is $\mathrm{ 2.3 \pm 4.5 \ (RMS) }$~mbar, within the
systematic uncertainty of the algorithm (Section~\ref{sec:systematics}).
This overall result supports the initial hypothesis that \DP~represents
primarily systematic offsets of pressure gauges and other sensors and,
once determined, would remain stable throughout a year of running.
 
\section{Resolution results}
\label{sec:MDTres}
 
The single-hit MDT tube spatial resolution, as defined in Equation~(\ref{eq:res}),
is measured in runs with the toroidal magnetic field on.
A requirement of \PT~$>$~20~\GeV\ for the muon transverse momentum
measured with the ID is applied to reduce the track curvature such
that track segments could be reliably fit with a linear function,
even in the chambers in a high magnetic field. For these tracks the
negligible effects of the curvature due to magnetic field are folded into
the systematic uncertainty.
\par
\begin{figure}[htpb]
\begin{minipage}[t]{0.48\linewidth}
\centering
\vspace{0pt}
\captionsetup{width=.95\linewidth}
\includegraphics[width=1.0\textwidth]{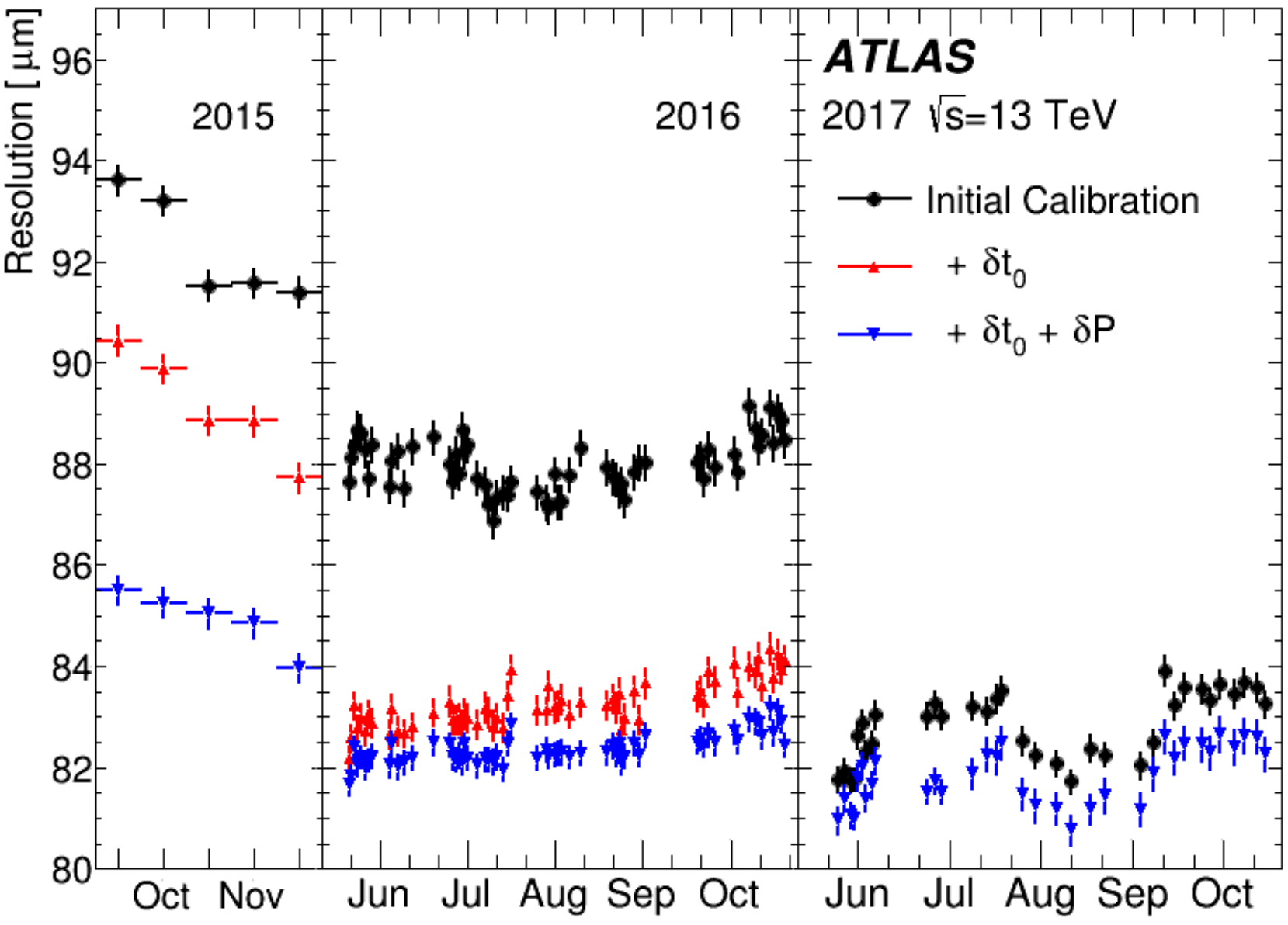}
\caption{Resolution averaged over chambers for runs in 2015
(left), 2016 (center) and 2017 (right) using default
calibrations, with \DTZ~and \DP~corrections.
The initial calibration data in the right panel already include the
\DTZ~corrections from 2016;  additional \DTZ~correction is not needed.}
\label{fig:ResBeforeAfterTPvsRun2015-2017}
\end{minipage}
\hfill
\begin{minipage}[t]{0.48\linewidth}
\centering
\vspace{0pt}
\captionsetup{width=.95\linewidth}
\includegraphics[width=1.0\textwidth]{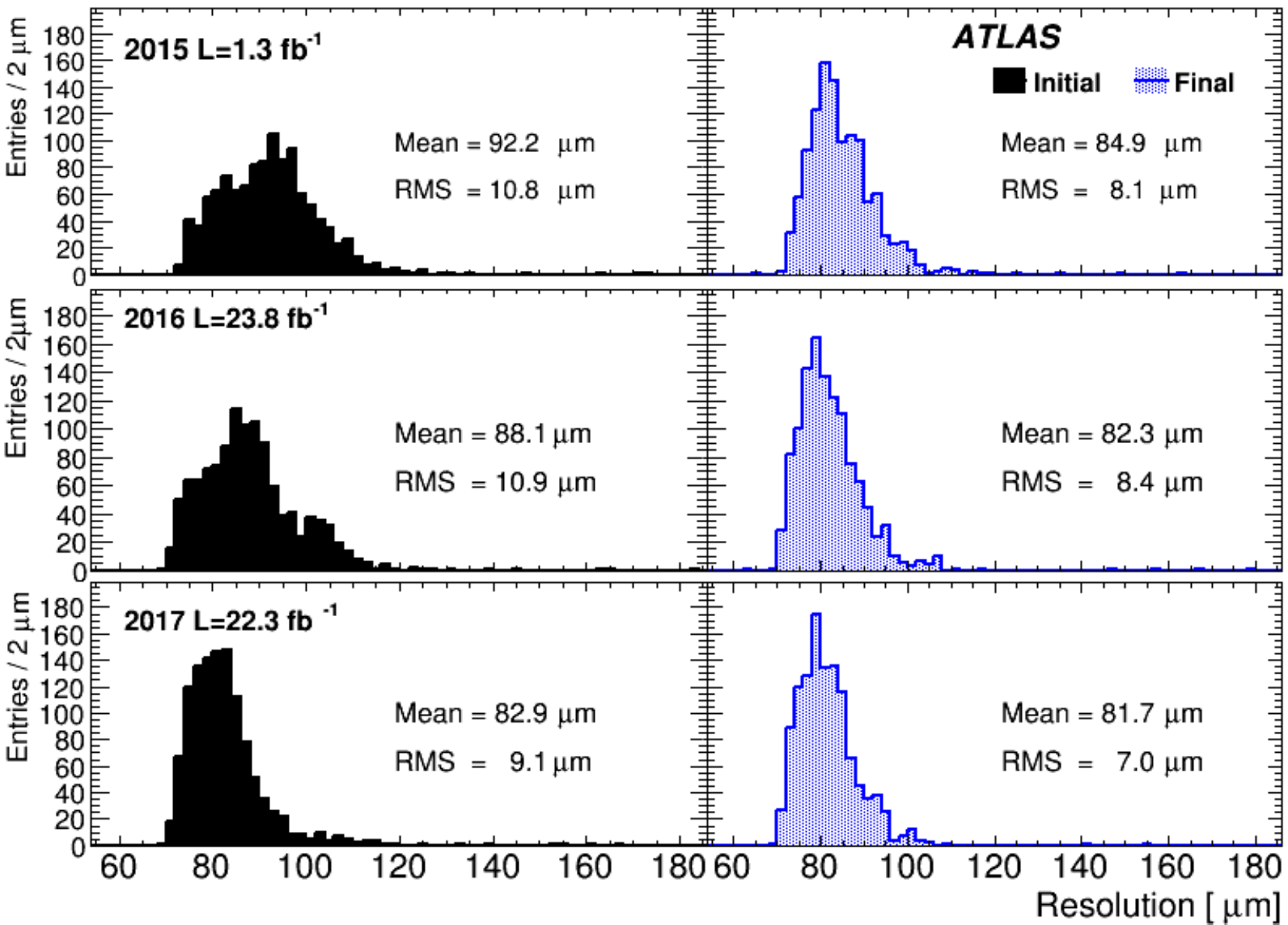}
\caption{Single-hit resolutions for each chamber.  Left: with
default calibrations (different conditions every year).
Right: with all corrections and most up-to-date \RT~functions.}
\label{fig:ResBeforeAfterTP2015-2017}
\end{minipage}
\end{figure}

The resolution obtained for a series of runs in each year
of Run~2 is reported in Figure~\ref{fig:ResBeforeAfterTPvsRun2015-2017}.
The full circles are calculated directly from fits to the
default reconstructed track segments using the standard \TZ~and
\RT~calibration constants.
The  triangles include the \DTZ~corrections, and the blue inverted
triangles also include \DP~corrections.
Significant improvements to the resolution in 2015 are the result of a
first implementation of the corrections, dominated by \DP.
These \DP~corrections were subsequently included in the standard production
of \RT~function calibration constants.
The application of the procedure in each year yields 1--2~$\mathrm \upmu$m
better resolution.
The \DTZ~corrections in 2015 and 2016 each lead to a 3--4~$\upmu$m
improvement in resolution.
From the beginning of 2017 the previous year's \DTZ~corrections are also
applied, but only the \DP~corrections from 2016 (not those from 2015).
However, they yield an improvement of only $\mathrm{1.2 \; \upmu m} $
as shown in the right panel of
Figure~\ref{fig:ResBeforeAfterTPvsRun2015-2017}.
\par
Both the best \DTZ and \DP~values are incorporated in the end-of-year data
reprocessing in order to provide the best calibrated data for
subsequent physics analyses.
 
Figure~\ref{fig:ResBeforeAfterTP2015-2017} summarizes the resolution averaged
over all runs in 2015, 2016, and 2017, for each MDT chamber.
The left panels show the initial resolutions obtained with the
standard calibrations.
The right panels show the resolutions after all \DTZ~and \DP~corrections
are implemented and using daily updated \RT~functions:
$\mathrm{ 84.9 \pm \; 0.2  \; \upmu}$m for 2015,
$\mathrm{ 82.3 \pm \; 0.2  \; \upmu}$m for 2016, and
$\mathrm{ 81.7 \pm \; 0.2  \; \upmu}$m for 2017.
Notably the improvements are  measured both by the decrease of
the mean value of the distribution and by the decrease of the RMS width,
indicating greater uniformity of the resolutions of the whole ensemble of MDT
chambers.
In 2015, the \DP~corrections were applied for the first time in Run~2,
and then incorporated into the default calibrations for 2016.
In 2016, the larger data set allowed a much higher fraction of the initial
\TZ~offsets to be determined for single tubes by the calibration centers,
resulting in a better overall initial resolution.
The \DTZ~corrections from 2016 were propagated into the default calibrations
for 2017.
For this reason, the incremental improvement in both the mean and the RMS
spread decreases with each year and the standard offline tube resolution
approaches the fully corrected result.
\par
\begin{figure}[htpb]
\begin{minipage}[t]{0.48\linewidth}
\centering
\vspace{0pt}
\captionsetup{width=.95\linewidth}
\includegraphics[width=\linewidth]{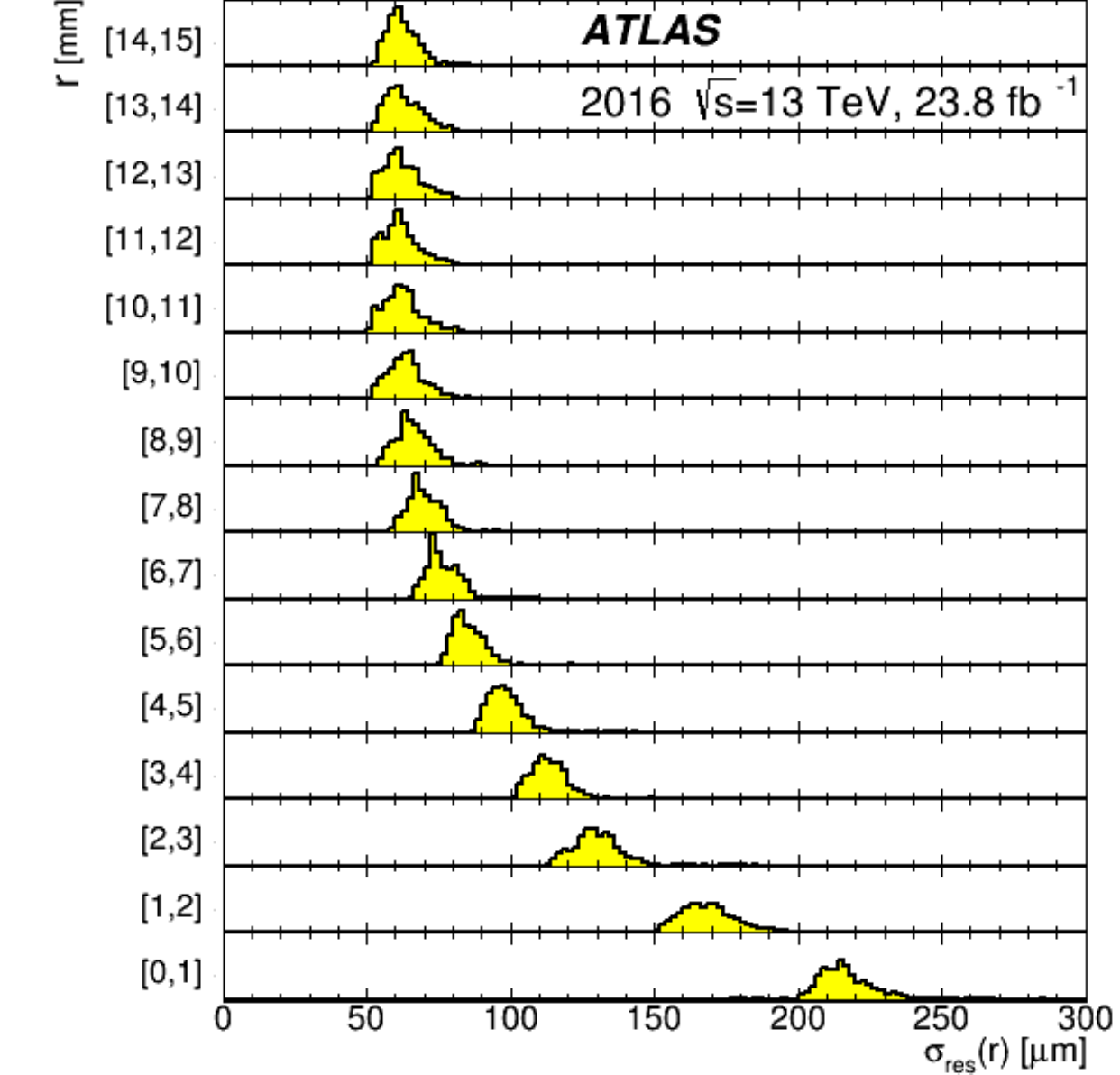}
\caption{Distribution of resolutions in 1~mm radial bins of drift
radius for all MDT chambers, averaged over all analysed runs from 2016.}
\label{fig:resallBon20GeV}
\end{minipage}
\hfill
\begin{minipage}[t]{0.48\linewidth}
\centering
\vspace{0pt}
\captionsetup{width=.95\linewidth}
\includegraphics[width=1.0\textwidth]{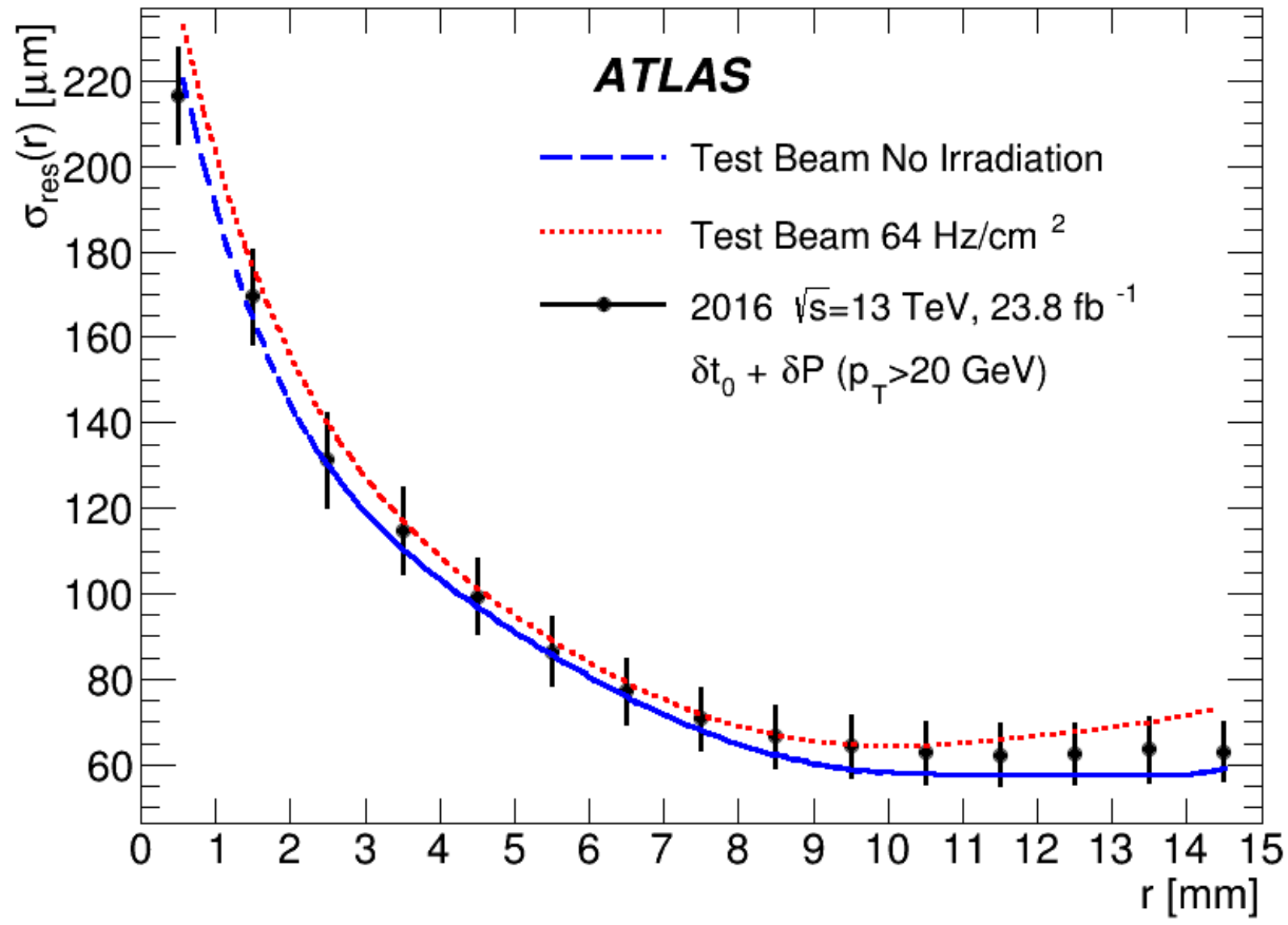}
\caption{Resolution as a function of drift radius averaged over all MDT
chambers and all analysed runs from 2016.  Muon test-beam resolutions
of an MDT chamber with background radiation from 0 to 64~Hz/cm$^2$
are also shown for comparison~\cite{Deile:2016ddq}.
Uncertainties are RMS widths of the resolution distribution for all
chambers.}
\label{fig:ResVsRad2016}
\end{minipage}
\end{figure}

The final MDT drift tube spatial resolution, with all corrections described
in previous sections included,
is shown as a function of drift radius in Figure~\ref{fig:resallBon20GeV}
and in Figure~\ref{fig:ResVsRad2016}.
In particular, Figure~\ref{fig:resallBon20GeV} shows for 2016 the run-averaged
resolution distributions $\mathrm{ \sigma_{res}}(r)$ in 1~mm radial bins,
where each histogram bin entry represents a single chamber.
This set of histograms therefore represents the full range of resolution
functions
obtained for the entire ensemble of MDT chambers.
The mean values of these distributions are plotted as a single resolution
function
in Figure~\ref{fig:ResVsRad2016}, where the uncertainties are the RMS widths
of the distributions for all chambers.  For comparison, the resolution curves of a single MDT chamber deployed in
a test beam~\cite{Deile:2016ddq}, with and without background irradiation,
are also displayed.
The 2016 chamber-weighted resolution averaged over all radial bins is
$\mathrm{ 86.4 \pm 0.2  \; \upmu}$m
and the average excluding the radial bin closest to the wire is
$\mathrm{ 79.7 \pm 0.2  \; \upmu}$m.
These two values bracket the final 2016 resolution averaged over all chambers,
presented in Figure~\ref{fig:ResBeforeAfterTP2015-2017}, of
$\mathrm{ 82.3 \pm 0.2  \; \upmu}$m.
 
\par
Specific high instantaneous luminosity collision runs are analysed in
order to measure the drift tube rates by counting all chamber hits
occurring within the MDT read-out time window of $\mathrm 1.2~\upmu$s
initiated by any muon trigger.
This total number of hits per time window is normalized to the total tube
area to obtain a rate per unit area.
These rates are consistently less than $\mathrm{ 10 \; Hz/cm^2}$,
except for one set of 16 endcap inner chambers at high pseudorapidity
whose rates range from 20 to 50 $\mathrm{ Hz/cm^2}$, and another set of 16
chambers nearest the beam-line where rates range from 60 to
140~$\mathrm{ Hz/cm^2}$.
At background rates above 100~$\mathrm{ Hz/cm^2}$, some loss in tracking
efficiency and resolution is expected due to increased read-out electronics
dead time and space charge in the gas, respectively.
Furthermore, an increase of the read-out electronics dead time was observed
in special high instantaneous luminosity runs at the end of Run~1, and
addressed in Run~2 by optimizing the data acquisition parameters.
The resolutions from the analysed runs recorded in 2017 are entirely
consistent with the results from 2016 shown here.
 
\section{Systematic uncertainties}
\label{sec:systematics}
The main sources of systematic uncertainty affecting the determination of the
MDT resolution are:
(1) the fit uncertainty associated with the range of the double-Gaussian fit;
(2) the use of linear fits for curved track segments;
(3) the precision of the \DTZ~and \DP~determinations; and
(4) the accuracy of Equation~\ref{eq:res} to describe the resolution.
They are described in  this section.
\par
A double-Gaussian function is fit in a range of $\mathrm {}\pm$500 $\upmu$m for
both the fit and hit residual distributions.
This fit function matches well the central region with a narrow component
and the tails with the wider component.
The fraction of hits included in the residual distributions exceeds 99\%
for most drift radii, and (95--97)\% for the first millimeter nearest the wire.
The reason for this additional loss is attributed to the minimum 1~mm
path length corresponding to the MDT front-end electronics
threshold of 23 primary ionization electrons~\cite{MDTelx:2008}.
For radial distances below 1~mm, the precision of the drift radius
determination deteriorates due to the small number of ion pairs, increasing
the fraction of large residuals.
A modification of the  fit range down to
${} \pm$400 $\upmu$m and up to ${}\pm$1~mm produces an overall
systematic variation in the resolution of $\mathrm 1 \; \upmu$m.

\par
The path of a low-\PT\ muon track is slightly deflected by the magnetic
field as it passes through a chamber.
For this reason, the tracks used in the analysis have a minimum transverse
momentum  of 20~\GeV, yielding an average \PT\ of 40~\GeV.
The systematic resolution broadening from using a linear fit is
determined from a Monte Carlo calculation using a worst-case 0.6~T
magnetic field and for the most prevalent MDT geometry:
a six-layer chamber with a 17~cm spacer frame separating the two multilayers.
The hits are smeared with the nominal $\mathrm 80 \; \upmu$m resolution,
linearly fit to generate residuals, and the resolution computed from
Equation~(\ref{eq:res}).
The average resolution for \PT\ = 40~\GeV\ muons is broadened by
about $\mathrm{ 2 \; \upmu}$m relative to the nominal value.
In addition, this  uncertainty from track curvature in the magnetic field
is also directly investigated by comparing the resolutions obtained with
and without the toroidal magnetic field.
The average resolution of all chambers with the toroid field
off is $\mathrm{ \sigma_{off}  = 80.1 \pm 0.2  \; \upmu}$m.
The average broadening of the measured resolution due to the toroidal field is
$\mathrm{ 1.6 \pm 0.3  \; \upmu}$m, consistent with the result from
Monte Carlo calculation.
\par
The step of the correction algorithm is 0.5~ns for \DTZ~and 5~mbar for \DP.
The minima are found separately for the hit and fit residual distributions,
and agree to within one step for more than 90\% of the chambers.
Therefore, the precision with which the parameters can be reliably established
is estimated to be half of the step size.
The uncertainty for each parameter is determined by altering the
optimal \DTZ~and \DP~values within ${} \pm$0.25~ns for \DTZ~and
${} \pm$2.5~mbar for \DP~and recalculating the resolution functions.
The average systematic uncertainty
in the correction parameter determination is $\mathrm 1 \; \upmu$m.
\par
A final, small systematic uncertainty derives from the accuracy to which
the resolution function of Equation~(\ref{eq:wg}) reproduces the true resolution.
This uncertainty, described in Section~\ref{sec:resolution}, is
$\mathrm 0.3 \; \upmu$m.
\par
The total systematic error of the single-hit spatial resolution measurement,
taken as the sum in quadrature of all contributions, is $\mathrm 2.2 \; \upmu$m,
as summarized in Table~\ref{table:sys},
and about 2.6\% of the overall average resolution.
\begin{table}[h]
\centering
\caption{Main resolution systematic uncertainties averaged over all chambers.}
\vspace{2mm}
\label{table:sys}
\begin{tabular}{| l | c |}
\hline
Systematic uncertainty         & Size [$\upmu$m] \\ \hline
Double-Gaussian fit range      & 1.0 \\
Straight-line track fit        & 1.6 \\
Optimal \DP~and \DTZ~values    & 1.0 \\
Resolution function accuracy   & 0.3 \\ \hline
Total                          & 2.2 \\ \hline
\end{tabular}
\end{table}

\FloatBarrier
 
\section {Conclusion}
 
This document describes a novel method to optimize the single-hit spatial
resolution of the muon MDT chambers of the ATLAS experiment during Run~2
of the LHC.
This method tunes the initial \TZ~derived from drift time spectra and
provides a correction parameter for \RT~functions produced by a gas monitor
MDT chamber. The extracted corrections were used in the 2015 and 2016 end-of-year data
reprocessing, and became fully integrated in the online calibrations
from  2017 onward, yielding results consistent with those presented here.
These two combined correction parameters are shown to be very stable
over a year, insensitive to increasing luminosity, variations in
environmental conditions, or the magnetic field.
The method applied to the 2017 data
achieves a mean single-hit spatial resolution of
$ \mathrm{ 81.7 \pm 0.2  \pm 2.2 \; (sys)\; \upmu}$m with an RMS spread
of $\mathrm{ 7\; \upmu }$m over all 1172 MDT chambers in the spectrometer.
The resolution approaches the design value of 80~$\upmu$m, consistent over
all MDT chambers, and stable across all run periods.

\section*{Acknowledgements}
 
% The next lines are included from the .//acknowledgements/Acknowledgements.tex input file
 
We thank CERN for the very successful operation of the LHC, as well as the
support staff from our institutions without whom ATLAS could not be
operated efficiently.
 
We acknowledge the support of ANPCyT, Argentina; YerPhI, Armenia; ARC, Australia; BMWFW and FWF, Austria; ANAS, Azerbaijan; SSTC, Belarus; CNPq and FAPESP, Brazil; NSERC, NRC and CFI, Canada; CERN; CONICYT, Chile; CAS, MOST and NSFC, China; COLCIENCIAS, Colombia; MSMT CR, MPO CR and VSC CR, Czech Republic; DNRF and DNSRC, Denmark; IN2P3-CNRS, CEA-DRF/IRFU, France; SRNSFG, Georgia; BMBF, HGF, and MPG, Germany; GSRT, Greece; RGC, Hong Kong SAR, China; ISF and Benoziyo Center, Israel; INFN, Italy; MEXT and JSPS, Japan; CNRST, Morocco; NWO, Netherlands; RCN, Norway; MNiSW and NCN, Poland; FCT, Portugal; MNE/IFA, Romania; MES of Russia and NRC KI, Russian Federation; JINR; MESTD, Serbia; MSSR, Slovakia; ARRS and MIZ\v{S}, Slovenia; DST/NRF, South Africa; MINECO, Spain; SRC and Wallenberg Foundation, Sweden; SERI, SNSF and Cantons of Bern and Geneva, Switzerland; MOST, Taiwan; TAEK, Turkey; STFC, United Kingdom; DOE and NSF, United States of America. In addition, individual groups and members have received support from BCKDF, CANARIE, CRC and Compute Canada, Canada; COST, ERC, ERDF, Horizon 2020, and Marie Sk{\l}odowska-Curie Actions, European Union; Investissements d' Avenir Labex and Idex, ANR, France; DFG and AvH Foundation, Germany; Herakleitos, Thales and Aristeia programmes co-financed by EU-ESF and the Greek NSRF, Greece; BSF-NSF and GIF, Israel; CERCA Programme Generalitat de Catalunya, Spain; The Royal Society and Leverhulme Trust, United Kingdom.
 
The crucial computing support from all WLCG partners is acknowledged gratefully, in particular from CERN, the ATLAS Tier-1 facilities at TRIUMF (Canada), NDGF (Denmark, Norway, Sweden), CC-IN2P3 (France), KIT/GridKA (Germany), INFN-CNAF (Italy), NL-T1 (Netherlands), PIC (Spain), ASGC (Taiwan), RAL (UK) and BNL (USA), the Tier-2 facilities worldwide and large non-WLCG resource providers. Major contributors of computing resources are listed in Ref.~\cite{ATL-GEN-PUB-2016-002}.
 
% End of text imported from the .//acknowledgements/Acknowledgements.tex input file

\clearpage

\DeclareFieldFormat{url}{\url{#1}}
\printbibliography
 
\clearpage
% ATLAS Collaboration author list
% Reference date of MDET-2018-01 is 2019-03-05
% Author list last updated on date 26-JUN-19
% Data extracted on 26-Jun-2019 for paper reference MDET-2018-01
% at 4:50pm
 
\begin{flushleft}
{\Large The ATLAS Collaboration}

\bigskip

G.~Aad$^\textrm{\scriptsize 101}$,    
B.~Abbott$^\textrm{\scriptsize 128}$,    
D.C.~Abbott$^\textrm{\scriptsize 102}$,    
O.~Abdinov$^\textrm{\scriptsize 13,*}$,    
A.~Abed~Abud$^\textrm{\scriptsize 70a,70b}$,    
K.~Abeling$^\textrm{\scriptsize 53}$,    
D.K.~Abhayasinghe$^\textrm{\scriptsize 93}$,    
S.H.~Abidi$^\textrm{\scriptsize 167}$,    
O.S.~AbouZeid$^\textrm{\scriptsize 40}$,    
N.L.~Abraham$^\textrm{\scriptsize 156}$
% ...
\bigskip
\\

$^{1}$Department of Physics, University of Adelaide, Adelaide; Australia.\\
% ...
$^{13}$Institute of Physics, Azerbaijan Academy of Sciences, Baku; Azerbaijan.\\
% ...
$^{40}$Niels Bohr Institute, University of Copenhagen, Copenhagen; Denmark.\\
% ...
$^{53}$II. Physikalisches Institut, Georg-August-Universit\"{a}t G\"ottingen, G\"ottingen; Germany.\\
% ...
$^{70}$$^{(a)}$INFN Sezione di Pavia;$^{(b)}$Dipartimento di Fisica, Universit\`a di Pavia, Pavia; Italy.\\
% ...
$^{93}$Department of Physics, Royal Holloway University of London, Egham; United Kingdom.\\
% ...
$^{101}$CPPM, Aix-Marseille Universit\'e, CNRS/IN2P3, Marseille; France.\\
$^{102}$Department of Physics, University of Massachusetts, Amherst MA; United States of America.\\
% ...
$^{128}$Homer L. Dodge Department of Physics and Astronomy, University of Oklahoma, Norman OK; United States of America.\\
% ...
$^{156}$Department of Physics and Astronomy, University of Sussex, Brighton; United Kingdom.\\
% ...
$^{167}$Department of Physics, University of Toronto, Toronto ON; Canada.\\
% ...
$^{a}$ Also at CERN, Geneva; Switzerland.\\
$^{b}$ Also at CPPM, Aix-Marseille Universit\'e, CNRS/IN2P3, Marseille; France.\\
% ...
$^{*}$ Deceased

\end{flushleft}

% Created with Glance <Atlas.Glance@cern.ch>

\end{document}